\documentclass[aps,pra,amsmath,amssymb,
reprint]{revtex4-1}
\usepackage{CJK}

\usepackage{graphicx}
\usepackage{dcolumn}
\usepackage{bm}

\usepackage[separate-uncertainty = true,multi-part-units=single]{siunitx} 
\usepackage{xcolor} 
\usepackage{ulem}

\begin{document}
\begin{CJK*}{GB}{}

\title{{Single-photon cooling in microwave magneto-mechanics}}
\author{D.~Zoepfl}
\thanks{equal contribution}
\email{david.zoepfl@uibk.ac.at}

\author{M.~L.~Juan}
\thanks{equal contribution}

\author{C.~M.~F.~Schneider}
\author{G.~Kirchmair}
\email{gerhard.kirchmair@uibk.ac.at}
\affiliation{Institute for Quantum Optics and Quantum Information of the Austrian Academy of Sciences, A-6020 Innsbruck, Austria
}
\affiliation{Institute for Experimental Physics, University of Innsbruck, A-6020 Innsbruck, Austria}

\date{\today}

\begin{abstract}
{Cavity optomechanics, where photons 
are coupled to mechanical motion, provides the tools to control mechanical motion near the fundamental quantum limits.} Reaching single-photon strong coupling would allow to prepare the mechanical resonator in non-Gaussian quantum states. Preparing  massive mechanical resonators in such states is of particular interest for testing the boundaries of quantum mechanics. This goal remains however challenging due to the small optomechanical couplings usually achieved with massive devices. Here we demonstrate a novel approach where a mechanical resonator is magnetically coupled to a microwave cavity. We measure a single-photon coupling of $g_0/2 \pi \sim \SI{3}{\kilo \hertz}$, an improvement of one order of magnitude over current microwave optomechanical systems. 
{At this coupling we measure a large single-photon cooperativity with $C_0 \gtrsim 10$, an important step to reach single-photon strong coupling. Such a strong interaction allows us to cool the massive mechanical resonator to a third of its steady state phonon population with less than two photons in the microwave cavity.} Beyond tests for quantum foundations, our approach is also well suited as a quantum sensor or a microwave to optical transducer.

\end{abstract}

\maketitle
{In recent years, cavity optomechanics has pushed the boundaries of quantum mechanics using micrometer-sized mechanical resonators}. Among the accomplishments were ground state cooling of mechanical motion~\cite{oconnell_quantum_2010, teufel_sideband_2011, chan_laser_2011}, measurement precision below the standard quantum limit~\cite{teufel_nanomechanical_2009, anetsberger_measuring_2010}, preparing mechanical resonators in non-classical states~\cite{wollman_quantum_2015, lecocq_quantum_2015, pirkkalainen_squeezing_2015, reed_faithful_2017} and entangling the mechanical state with the optical field~\cite{palomaki_entangling_2013, riedinger_remote_2018, ockeloen-korppi_stabilized_2018}. 
{In this context, an important parameter is the interaction strength between the mechanical and photonic modes: the optomechanical coupling. Recently, the ultrastrong coupling regime was reached where this coupling, enhanced by the photons in the cavity, exceeds both decay rates (cavity and mechanics) and is comparable to the mechanical frequency~\cite{peterson_ultrastrong_2019-1}. However, to achieve a non-linear optomechanical interaction, the coupling has to be further increased in order to reach the single-photon strong coupling regime}. This regime, where the single-photon coupling strength, $g_0$, exceeds both the linewidth of the cavity, $\kappa$, and the linewidth of the mechanical resonator, $\Gamma_m$, ($g_0 > \kappa, g_0 > \Gamma_m$) opens the door to prepare quantum superposition states in a mechanical resonator~\cite{aspelmeyer_cavity_2014}. {Large couplings can be achieved by using resonators with small masses~\cite{brennecke_cavity_2008,murch_observation_2008} or by replacing the cavity by a qubit~\cite{pirkkalainen_cavity_2015,chu_creation_2018,viennot_phonon-number-sensitive_2018,delsing_2019_2019,bera_large_2020}, although in the later case it is not possible to benefit from a photon enhanced coupling.
Reaching the single-photon strong coupling regime with mechanical resonators having a large mass and long coherence time is of particular interest to investigate the classical to quantum transition~\cite{arndt_testing_2014}. This remains challenging since the coupling depends directly on the zero-point fluctuation of the resonator: massive resonators generally exhibit much smaller couplings~\cite{aspelmeyer_cavity_2014}}.

A promising candidate for achieving single-photon strong coupling is microwave optomechanics, as it provides high quality cavities with much lower frequencies and is particularly well adapted to cryogenic operation~\cite{regal_cavity_2011}. {To date, the favoured approach for cavity optomechanics relies on a mechanically compliant element which modulates the capacitance of a microwave cavity}. Ultimately bounded by the capacitor gap and the zero-point fluctuation amplitude of the resonator, state-of-the-art devices have reached couplings of a few hundreds of Hertz~\cite{aspelmeyer_cavity_2014}. Achieving single-photon strong coupling presents extreme technological challenges in order to either increase the coupling strength $g_0$ or decrease the cavity linewidth substantially. 
{An important step towards this regime is to achieve a large single-photon cooperativity~\cite{aspelmeyer_cavity_2014}, $C_0 = 4 g_0^2 / \kappa \Gamma_m$. {For $C_0 > 1$, the back-action on the mechanical resonator from a single cavity photon is sufficient to enable cooling~\cite{yuan_large_2015}.}} This regime was recently achieved in the optical regime with massive resonators~\cite{guo_feedback_2019}, but remains challenging in the microwave regime due to the smaller coupling strengths.
\begin{figure*}[!t]
    \centering
    \includegraphics[width = 17.2cm]{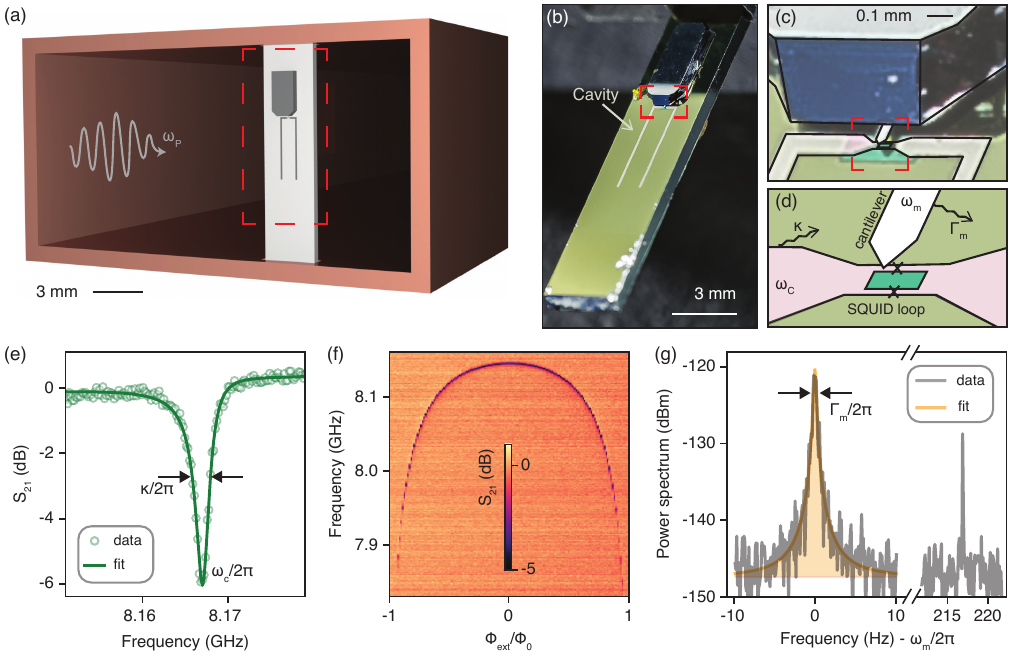}
    \caption{General setup and device characterisation. (a) Rectangular waveguide used to probe the microwave cavity, a U-shaped microstrip resonator on a Silicon substrate. (b) Photograph of the device showing the microwave cavity and cantilever chip. (c) Close-up of the cavity SQUID and the mechanical cantilever. (d) Sketch of the SQUID loop and cantilever. (e) {Transmission measurement through the waveguide (notch configuration) showing the microwave cavity response. A fit to the data gives a frequency $\omega_c/2 \pi = \SI{8.167}{GHz}$ and a linewidth $\kappa/2 \pi = \SI{2.8}{MHz}$}. (f) Change of the microwave cavity frequency, obtained by transmission measurements, as a function of the applied external magnetic flux. (g)
    {Thermal noise power spectrum of the cantilever at \SI{100}{mK} after amplification and homodyne down mixing when probing the cavity with a weak resonant microwave tone of \SI{-54.5}{dBm} fridge input power. 
    A fit to the power spectrum gives $\omega_m/2 \pi = \SI{274383.13(3)}{Hz}$ and $\Gamma_m/2 \pi = \SI{0.3(1)}{Hz}$, the area below the curve (coloured) corresponds to the motional energy of the cantilever mode. The sharp peak detuned $\sim \SI{215}{Hz}$ away from the mechanical frequency is the calibration peak~\cite{noauthor_see_nodate}.}}
    \label{fig:setup}
\end{figure*}

{Here we report on reaching a coupling strength in the \SI{}{\kilo\hertz} range, allowing us to demonstrate a single-photon cooperativity exceeding unity between a microwave cavity and a massive mechanical resonator.} To increase the coupling, we propose an alternative to most microwave experiments by magnetically coupling the mechanical resonator to the cavity, an approach which gained attention recently~\cite{via_strong_2015, rodrigues_coupling_2019, wang_preparing_2014, schmidt_sideband-resolved_2019}. Concretely, our mechanical resonator is a single clamped beam - a cantilever - with a magnetic tip. In order to mediate the optomechanical interaction, we integrated a superconducting quantum interference device (SQUID) in a U-shaped microstrip resonator~\cite{zoepfl_characterization_2017} to effectively obtain a microwave cavity sensitive to magnetic flux. The single-photon coupling strength, $g_0$, is given by the change of the cavity frequency, $\omega_c$, induced by the zero-point fluctuation, $x_{ZPM}$, of the mechanical resonator:
\begin{equation}
    g_0 = \frac{\partial \omega_c}{\partial x} x_{ZPM} = \frac{\partial \omega_c}{\partial \phi_{\text{ext}}} \times \frac{\partial \phi_{\text{ext}}}{\partial x} x_{ZPM}.
    \label{equ:g0}
\end{equation}
As $\partial \omega_c / \partial x$ is not directly accessible it is more convenient to express the coupling in terms of external magnetic flux $\phi_{\text{ext}}$:~The second part, $\partial \phi_{\text{ext}} / \partial x \times x_{ZPM}$, gives the flux change induced by a zero-point motion of the mechanical cantilever. {The first part describes the cavity frequency dependence on the flux through the SQUID loop hence providing a direct control of the coupling strength.}

Our experiment is mounted to the base plate of a dilution refrigerator~\cite{noauthor_see_nodate}. The microwave cavity is placed in a rectangular waveguide in order to provide a lossless microwave environment and control its coupling to the microwave probe tone travelling through the waveguide (see Fig.~\ref{fig:setup}(a)). We use the fundamental $\lambda/2$ mode with a current maximum at the centre, the position of the SQUID loop (see Fig.~\ref{fig:setup}). For the mechanical resonator we use a commercial atomic force microscopy cantilever having a nominal room temperature frequency of \SI{350}{kHz} and a mass of a few tens of nanograms~\cite{noauthor_see_nodate}. To mediate the magnetic coupling to the cavity, we functionalised its tip with a strong micromagnet (NdFeB) and completed the sample by placing the cantilever \SI{20(1)}{\micro\meter} above the SQUID, Fig.~\ref{fig:setup}(c) and (d).

\begin{figure}[t!]
    \centering
    \includegraphics[width = 8.6cm]{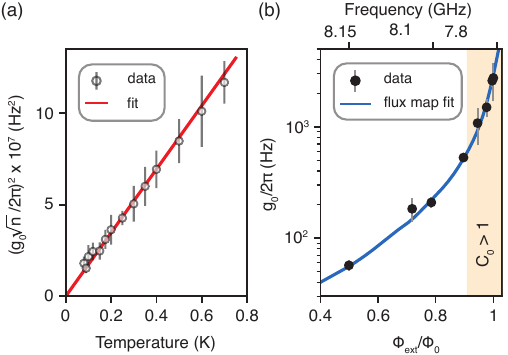}
    \caption{Temperature ramp and coupling strength dependence with the flux bias point. 
    {(a) Measurement of $g_0 \sqrt{n}$ with increasing cryostat temperature at a fixed sensitivity and fridge input power of \SI{-54.5}{dBm}~\cite{noauthor_see_nodate}. We verify that the cantilever is thermalised with the cryostat temperature by fitting it assuming $\left<n\right> = \left<n^{\text{th}}\right>$, obtaining a coupling $g_0/2 \pi = \SI{48(1)}{Hz}$.} (b) Measurement of the bare coupling strength for different flux bias points. The solid line is the sensitivity predicted from the slope of the flux map (Fig.~\ref{fig:setup}(f)). 
    {In the shaded region the coupling is sufficient to reach $C_0>1$.} (a) and (b), the error bar denotes the standard deviation of multiple measurements~\cite{noauthor_see_nodate}.}
    \label{fig:coupling_meas}
\end{figure}
The microwave cavity response is obtained via transmission measurements through the waveguide, Fig.~\ref{fig:setup}(e)~\cite{noauthor_see_nodate}. {Fitting the model for a resonator in notch configuration~\cite{probst_efficient_2015, khalil_analysis_2012} to our line shape gives a frequency $\omega_c / 2 \pi = \SI{8.167}{GHz}$ and a linewidth $\kappa / 2 \pi = \SI{2.8}{MHz}$, where coupling to the waveguide and internal losses contribute equally to the linewidth with $\kappa_c / 2 \pi \simeq \kappa_I / 2 \pi \simeq \SI{1.4}{MHz}$.} The internal losses set a lower bound on the total linewidth. To control the cavity frequency, an external magnetic field is applied through the SQUID loop by using coils, Fig.~\ref{fig:setup}(f). The slope of the flux map gives the sensitivity to magnetic fields, $\partial \omega_c / \partial \phi_{\text{ext}}$, which directly sets the coupling (equation~\ref{equ:g0}). The mechanical resonator modulates the response of the microwave cavity at its frequency $\omega_m$. We use a microwave probe tone close to the cavity resonance which, in the bad cavity limit $\kappa \gg \omega_m$, is amplitude modulated at the mechanical frequency. By performing a homodyne measurement we directly obtain the thermal noise power spectrum from the cantilever, Fig.~\ref{fig:setup}(g). A fit with a damped harmonic resonator model gives a mechanical frequency $\omega_m/2 \pi = \SI{274383.13(3)}{\hertz}$ and a linewidth $\Gamma_m/2 \pi =  \SI{0.3(1)}{Hz}$.


To extract the coupling between the cavity and the mechanical cantilever, we measure the thermal noise power spectrum which depends on the bare coupling enhanced by the phonon number: $g_0 \sqrt{n}$, but also on the transduction from the microwave cavity. To gain direct access to this transduction we apply a frequency modulation to the microwave probe tone~\cite{gorodetksy_determination_2010, zhou_slowing_2013}. 
{Using such a calibration tone, Fig.~\ref{fig:setup}(g), we get instant access to the transduction of the microwave cavity at the measurement point and directly obtain the value of $g_0 \sqrt{n}$ from the power spectrum~\cite{noauthor_see_nodate}.} In addition, extracting the bare coupling $g_0$ requires knowledge of the phonon number. In the absence of optomechanical back-action and excessive vibrations, we expect the mechanical mode to be thermalised with the cyrostat, $\left<n^{\text{th}}\right> = 1/(e^{\hbar \omega_m / k_b T} -1 ) \simeq k_B T / (\hbar \omega_m)$, where $k_B$ is the Boltzmann constant, $T$ the temperature and $\hbar$ the reduced Planck constant.

In order to verify that the mechanical mode is thermalised, we increased the temperature of our cryostat from \SI{80}{mK} to \SI{700}{mK}, Fig.~\ref{fig:coupling_meas}(a), and measured $g_0 \sqrt{n}$. Keeping $g_0$ constant, we expect an increase in $g_0 \sqrt{n}$ due to an increasing phonon population with the cryostat temperature. By fitting the data assuming $\left<n\right> = \left<n^{\text{th}}\right>$, we extract a bare coupling of $g_0/2 \pi = \SI{48(1)}{Hz}$. To avoid any optomechanical back-action, we chose a point of weak coupling along with a moderate microwave probe tone of \SI{-54.5}{dBm}~\cite{noauthor_see_nodate}, as the photon enhanced coupling has to be considered for the back-action. This assumption is verified by ensuring $g_0 \sqrt{n}$ is constant while varying the input power~\cite{noauthor_see_nodate}. All the following measurements were done at \SI{100}{mK}.

Next, we demonstrate the control of the coupling strength, $g_0 \propto \partial \omega_c / \partial \phi_{\text{ext}}$, by changing the external flux bias. To avoid any back-action on the cantilever, we reduced the power in the microwave cavity according to the increasing flux sensitivity~\cite{noauthor_see_nodate}. The measured coupling strength in dependence with the flux bias point is shown in Fig.~\ref{fig:coupling_meas}(b). The solid line depicts the sensitivity, which is extracted from the derivative of the flux map, Fig. \ref{fig:setup}(f), where a fit to the data provides the flux change per phonon $\partial \phi_{\text{ext}} / \partial x \times x_{ZPM}=\SI{1.60(5)}{\micro\phi_0}$. The main limitation for measuring higher couplings, in addition to increased flux instability during the 10 minutes measurement time, is the much lower signal as we reduced the incident probe power. 
{For the highest couplings measured, $g_0/2 \pi \sim \SI{3}{\kilo\hertz}$, we achieve a large single-photon cooperativity of $C_0 \gtrsim 10$.}
\begin{figure}[t]
    \centering
    \includegraphics[width = 8.6cm]{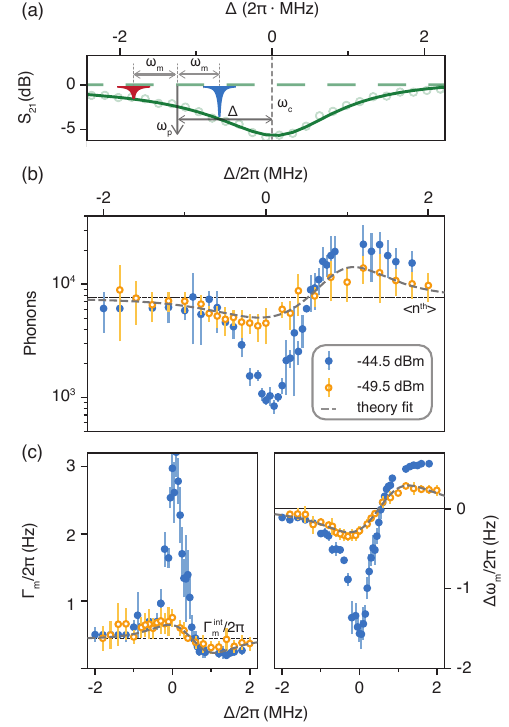}
    \caption{Back-action measurements for small coupling, $g_0/2 \pi$ = \SI{57(7)}{\hertz}. (a) Illustration of the cavity cooling mechanism for a pump tone $\omega_p$, red-detuned from the cavity, $\Delta = \omega_p - \omega_c < 0$. In blue and red are the anti-Stokes and Stokes scattering processes respectively. (b) Phonon number in the mechanical cantilever against detuning of the pump for two different powers. The dashed line shows the thermal phonon number for the mode $\left<n^{\text{th}}\right>\simeq 7600$. (c) Change of linewidth and mechanical frequency against detuning of the pump.
    {A fit of the phonon number with detuning for the lower power measurement gives a maximum photon number of $\SI{186(12)}{}$ and $g_0/2 \pi = \SI{57(1)}{Hz}$. The predictions from the theoretical model using the extracted parameters are plotted in (b) and (c).} (b) and (c) the error bar denotes the propagated fit error of multiple measurements~\cite{noauthor_see_nodate}.}
    \label{fig:backaction_measWeak}
\end{figure}

While previous measurements were obtained with low enough input power to avoid back-action, we discuss in the following the possibility to cool the mechanical mode. By driving the cavity red-detuned ($\omega_p < \omega_c$) inelastic anti-Stokes scattering is favoured leading to cooling of the mechanical motion~\cite{aspelmeyer_cavity_2014} (Fig.~\ref{fig:backaction_measWeak}(a)). In addition, such back-action is accompanied by a broadening of the mechanical linewidth and a frequency shift. Conversely, pumping blue-detuned leads to heating of the mechanics and a decrease of the linewidth. Dynamical instability of the mechanical mode is reached when the linewidth approaches zero~\cite{marquardt_quantum_2007}.

First, we demonstrate cavity cooling by operating at a low coupling, $g_0/2 \pi = \SI{57(7)}{Hz}$ (Fig.~\ref{fig:coupling_meas}(b)). 
{For low power (open symbols in Fig. 3), we fit the back-action measurements using the theory for cavity-assisted cooling~\cite{marquardt_quantum_2007,noauthor_see_nodate,safavi-naeini_laser_2013}, obtaining an independent measurement of the maximum photon number of $\SI{186(12)}{}$ and the coupling $g_0/2 \pi = \SI{57(1)}{Hz}$. As expected, the change of phonon number is accompanied by a linewidth and frequency change (Fig.~\ref{fig:backaction_measWeak}(c)). We note that we also included a frequency offset to the fit to accommodate the impedance mismatch of the cavity with the waveguide~\cite{noauthor_see_nodate}. For increasing input power the back-action increases, allowing us to achieve a nearly 8 fold decrease from the thermal phonon occupation $\left<n^{\text{th}}\right>\simeq 7600$ to $\left<n\right> = \SI{970(130)}{}$,  Fig.~\ref{fig:backaction_measWeak}(b). Since we are in the bad cavity regime, the theoretical limit is given by $\left< n ^{\text{min}} \right> = (\kappa / 4 \omega_m)^2 \simeq 6.5$~\cite{marquardt_quantum_2007}. Practically, we were limited by the non-linearity of the microwave cavity arising from the Josephson Junctions. This effect prevents us from fitting the data for higher input power, but also sets an upper bound on the cavity photons of around \SI{1200}{} before it becomes bistable~\cite{noauthor_see_nodate,nation_quantum_2008}. The impact of the non-linearity, while it constitutes an interesting element of study on its own~\cite{nation_quantum_2008}, could be mitigated by further improving the cavity, for example by using a SQUID array.}
\begin{figure}[t]
    \centering
    \includegraphics[width = 8.6cm]{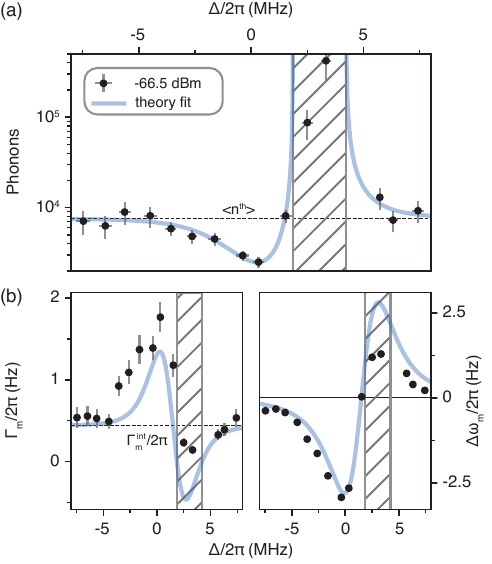}
    \caption{{Back-action measurements in the large coupling regime, where $C_0 \gtrsim 10$. (a) Phonon number against detuning. A fit provides the maximum photon number in the microwave cavity, \SI{2.1(4)}, and $g_0/2 \pi = \SI{2.38(6)}{kHz}$. (b) Linewidth and frequency shift against detuning. We plot the theoretical predictions with the fit parameters obtained in panel (a). (a) and (b), the hashed area marks the region of dynamical instability.} The $y$-error bar denotes the propagated fit error of multiple measurements, the $x$-error arises from our grouping method we apply to this data~\cite{noauthor_see_nodate}.}
    \label{fig:backaction_measStrong}
\end{figure}

{To demonstrate cooling with a few photons, we work at a more flux sensitive point where we expect a coupling $g_0/2 \pi = \SI{2.3(3)}{\kilo\hertz}$ (Fig.~\ref{fig:coupling_meas}(b)) and a large cooperativity $C_0 \gtrsim 10$.} By using an input power of \SI{-66.5}{dBm}, for which we expect an average occupation of around a single photon in the cavity, we clearly demonstrate cooling and heating of the cantilever mode, Fig.~\ref{fig:backaction_measStrong}. {By fitting the experimental data~\cite{marquardt_quantum_2007,noauthor_see_nodate,safavi-naeini_laser_2013} we extract a maximum photon number of \SI{2.1(4)}, and a coupling of $g_0/2 \pi = \SI{2.38(6)}{kHz}$. In terms of cooling, owing to the large cooperativity, we reached $\left<n\right> = \SI{2430(310)}{}$, corresponding to a cooling factor $\sim 3$ for a nanogram-scaled mechanical resonator. For the corresponding detuning, we extract a cavity population of only 1.4 photons~\cite{noauthor_see_nodate}. We note that the mechanical linewidth is significantly larger than the theoretical prediction, an effect that we attribute to the high sensitivity to flux noise~\cite{noauthor_see_nodate}. By increasing the incident power until the non-linearity was too severe, we reached $\left<n\right> \simeq \SI{150}{}$ with $\sim 20$ photons in average in the cavity.}


To conclude, the novel approach for microwave optomechanics we demonstrate in this Letter relies on simple elements, namely a $\lambda/2$ superconducting resonator with an integrated SQUID for the cavity and a commercial cantilever for the mechanical resonator, providing an optomechanical $g_0$ in the \SI{}{\kilo\hertz} range. 
{In the context of cavity optomechanics, this constitutes an improvement of one order of magnitude over couplings achieved in the microwave regime. Owing to this strong interaction, we achieved large single-photon cooperativities of $C_0 \gtrsim 10$ and demonstrated the cooling of the mechanical mode to a third of its thermal population with less than two photons in the cavity.} Furthermore, owing to the 3D architecture of this approach, it offers numerous opportunities to significantly improve the optomechanical coupling as well as decreasing the microwave cavity linewidth, clearly paving the way to enter single-photon strong coupling. {This would, most notably, facilitate preparing the massive mechanical resonator in non-Gaussian states which can be used to perform fundamental tests on quantum mechanics.} In addition, our approach can be used in more practical applications such as force sensing~\cite{mamin_sub-attonewton_2001} and microwave-to-optics transduction~\cite{andrews_bidirectional_2014}.


\textbf{Acknowledgment:} We want to thank Hans Huebl for fruitful discussion. Further we thank our in-house mechanical workshop. DZ and MLJ are funded by the European Union$'$s Horizon 2020 research and innovation program under grant agreement No. 736943. CMFS is supported by the Austrian Science Fund FWF within the DK-ALM (W1259-N27).


\nocite{yuan_silicon_2015,agrawal_optical_1979}


%

\clearpage
\onecolumngrid

\begin{center}
\textbf{\large Supplemental information: Single-photon cooling in microwave magneto-mechanics}
\end{center}

\setcounter{equation}{0}
\setcounter{figure}{0}
\setcounter{table}{0}
\setcounter{page}{1}
\makeatletter
\renewcommand{\theequation}{S\arabic{equation}}
\renewcommand{\thefigure}{S\arabic{figure}}

\vspace{0.5 in}

\section{Sample preparation}

\begin{figure*}[h]
    \centering
    \includegraphics[width = 8.6cm]{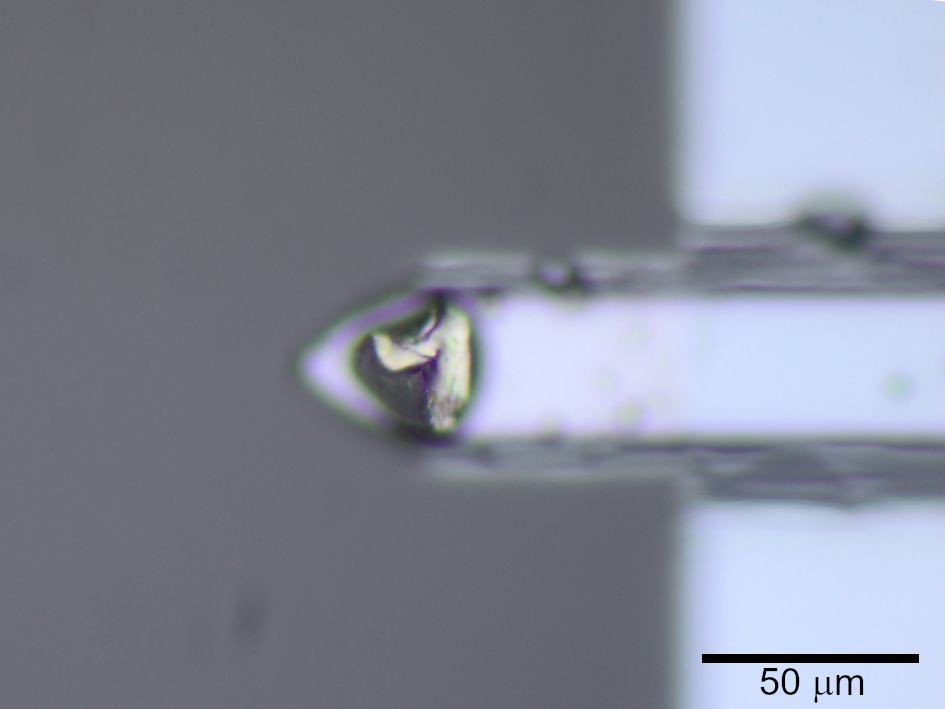}
    \caption{Photograph of the cantilever with the magnet before assembly to the microwave cavity.}
    \label{figS:pic_canti}
\end{figure*}

Our sample consists of two separate chips: one with the microwave cavity and one with the mechanical cantilever. The microwave cavity, custom made by Star Cryoelectronics, consists of a U-shaped Niobium film on a Silicon substrate with a total length of \SI{8}{mm}. The SQUID loop has dimensions of $\SI{60}{\micro \metre} \times \SI{20}{\micro \metre}$. The Josephson Junction are of circular shape with a \SI{4}{\micro \metre} diameter. To pattern the Josephson Junctions, a Nb/Al/Nb trilayer process is used. For the mechanical resonator we use commercial all-in-one tipless atomic force microscopy cantilevers from BudgetSensors with dimensions of \SI{100}{\micro\metre} x \SI{50}{\micro\metre} x \SI{3}{\micro\metre}. Each cantilever is functionalized with a NdFeB strong micromagnet (Fig.~\ref{figS:pic_canti}) and then magnetized it with a \SI{2}{\tesla} magnet in order to present a magnetic moment normal to the SQUID loop to maximise the coupling. The complete sample is assembled under an optical microscope by aligning the tip of the cantilever with the SQUID loop of the microwave cavity and secure it using GE varnish. The cantilever used in this Letter was placed \SI{20(1)}{\micro\meter} away from the microwave cavity and was prepared with a micromagnet with a pyramidal shape with a triangular base with sides of approximately \SI{28}{\micro\meter} and a height of \SI{8}{\micro\meter}. To finish the preparation, the sample is inserted in a designated slot of our rectangular waveguide. We estimate the mass of the cantilever using two different methods. In one method we estimate the mass using the volume and known densities, where we estimate the cantilever and magnet size via the microscope picture, which gives a mass of \SI{28}{ng}. Using the second method, we evaluate the effective mass using the known frequency and the nominal force constant provided by the manufacturer of the cantilever, where we evaluate \SI{14}{ng}.

\section{Full measurement setup}
\begin{figure*}[t]
    \centering
    \includegraphics[width = 17.2cm]{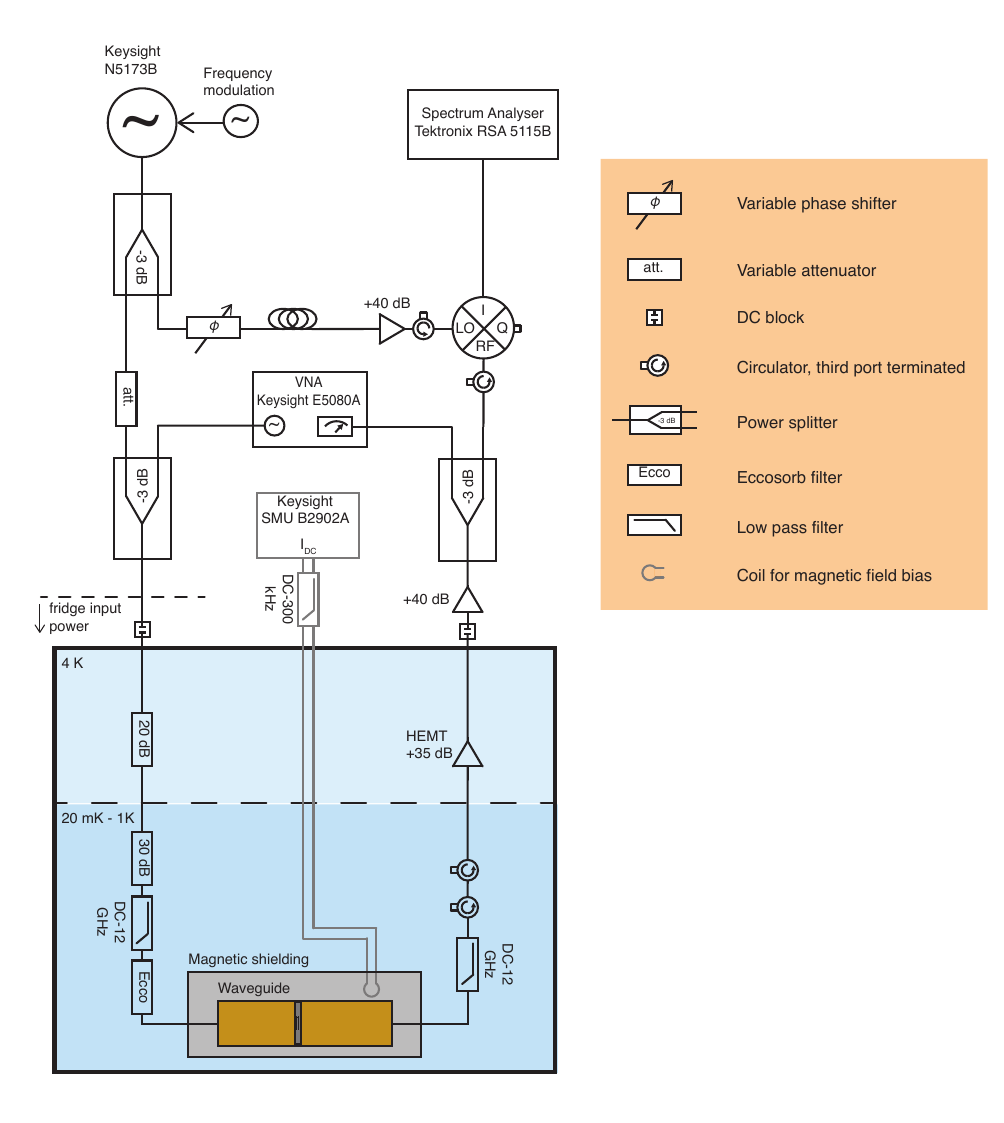}
    \caption{Full measurement setup. Details in the main text.}
    \label{figS:full_setup}
\end{figure*}

Here we describe the full measurement setup, illustrated in Fig.~\ref{figS:full_setup}. The setup allows to acquire a power spectrum using a fixed frequency probe tone and to do a vector network analyser (VNA) measurement. To enable this, we combine the signals going to the cryostat and split them on the output using conventional power splitters.

When acquiring a power spectrum, we perform a homodyne down-mixing of the measurement signal using an IQ mixer. The calibration routine (see section~\ref{secS:cali}) requires to use the same frequency modulated probe tone for the local oscillator (LO) port of the mixer and the experiment going to the radio frequency (RF) port of the mixer. Furthermore the calibration routine requires equal electrical lengths in both lines, which we fine tuned using a variable phase shifter. Due to the cable length, amplification of the signal before the LO port is necessary to provide sufficient power to the mixer. For all measurements we kept the power output of the frequency generator fixed to \SI{-7}{dBm} and control the power sent to the cryostat by a variable attenuator. For most measurements we measure using the in-phase (I) port of the mixer. When operating off resonance, for some detuning the I quadrature vanishes, in this case we used the quadrature (Q) port.

By adding the attenuation of all the independent components on the cryostat input line, we get an approximate input attenuation, which allows us to estimate the incident power on the microwave cavity and hence the photon number. The input power we refer to throughout the manuscript is the input power at the top of the cryostat (Fig.~\ref{figS:full_setup}). We estimate the additional attenuation from the top of the fridge to the sample with \SI{63}{dB}.
In addition, to control the flux bias point, we use a coil wound around the waveguide body, driven by a current source. The waveguide is made from copper and sits in a double layer mu-metal shield surrounded by a copper shield.

\section{Calibration method used to determine the optomechanical coupling strength} \label{secS:cali}
We follow the calibration method described in~\cite{Sgorodetksy_determination_2010,zhou_slowing_2013}. The concept relies on frequency modulating the probe tone at a frequency similar to the mechanical frequency ($\omega_{mod} \approx \omega_{m}$). Due to this modulation, an additional sideband appears on the power spectrum which carries information on the transduction from microwave cavity. The magneto-mechanical coupling strength $g_0$ is then obtained as:
\begin{equation}
    g_0^2 \approx \frac{1}{2 \left<n\right>} \frac{\phi_0^2 \omega_{mod}^2}{2} \frac{S_{II}^{\text{meas}} (\omega_m) \Gamma_m/4}{S_{II}^{\text{meas}} (\omega_{mod}) \text{ENBW}}.
    \label{equS:g_0_calibration}
\end{equation}
Here $\left<n\right>$ is the average phonon number, $S_{II}$ are the intensity-intensity fluctuations in the power spectral density at $\omega_m$ and $\omega_{mod}$, ENBW is the instrument effective noise bandwidth and $\phi_0$ is the modulation index of the frequency modulation. It is defined as $\omega_{Dev} / \omega_{mod}$, where $\omega_{Dev}$ is the development of the modulation (over which frequency range the probe frequency is modulated). We note that this expression is valid if the transduction at $\omega_{mod}$ is similar to the transduction at $\omega_{m}$. To extract the linewidth $\Gamma_m$, we fit the mechanical peak in the homodyne spectrum with the model for a damped harmonic oscillator~\cite{Szhou_slowing_2013}:
\begin{equation}
    S_{\omega \omega} (\omega) \approx g_0^2 \left<n\right> \frac{2 \omega_m}{\hbar} \frac{2 \Gamma_m}{(\omega^2 - \omega_m^2)^2 + \Gamma_m^2 \omega^2},
    \label{equS:PSD_HO}
\end{equation}
with the approximation being valid for large occupation numbers, $\left<n\right> \gg 1$. 

A key aspect for the calibration method to work, is that it requires an equal electrical length of the cable through the fridge to the RF port of the mixer and of the cable to the LO port. With this the phase of the frequency modulation in the LO and RF port of the mixer are identical and only the additional effect from the transduction of the microwave cavity is observed in the spectrum, which constitutes our calibration tone. We vary the development, $\omega_{Dev}$, accordingly to the input power to keep the absolute signal of the calibration tone at a similar level. For the temperature ramp we used \SI{2}{kHz}, while for the cooling traces with large coupling (and low photon number) we used \SI{200}{kHz}.

\section{Full characterisation of the microwave cavity}
Here we present the full characterisation of the microwave cavity using the circle fit routine~\cite{Sprobst_efficient_2015}. The U-shaped microwave cavity in the waveguide constitutes a resonator in notch configuration. We measure the cavity in transmission, which is described by the following model~\cite{Sprobst_efficient_2015}:
\begin{equation}
     S_{21} (\omega) =  1 - \frac{Q_l/|Q_c| e^{i \phi_0}}{1 + 2 i Q_l \frac{\omega - \omega_c}{\omega_c}}.
     \label{equS:S21_ideal}
\end{equation}
Here $Q_l$ is the total quality factor, $\omega_c$ is the cavity frequency and $Q_c$ is the coupling quality factor. In addition, $\phi_0$ accounts for an impedance mismatch in the microwave transmission line before and after the resonator, which makes $Q_c$ a complex number ($Q_c = |Q_c| e^{- i \phi_0}$). The real part of the coupling quality factor determines the decay rate of the resonator to the transmission line, in our case the emission to the waveguide. The physical quantity is the decay rate, $\kappa$, which is inversely proportional to the quality factor~\cite{Skhalil_analysis_2012} and therefore the real part is found as: $1/Q_c^{Re} = \text{Re} (1 / Q_c) = \cos{\phi_0} / |Q_c|$.  Knowing $Q_c^{Re}$ and $Q_l$, the internal quality factor can be obtained, as $1/Q_l = 1/Q_c^{Re} + 1/Q_{\text{int}}$~\cite{Skhalil_analysis_2012}. Plotting the imaginary versus the real part of $S_{21}$ forms a circle in the complex plane.

Equation~\ref{equS:S21_ideal} represents an isolated resonator, not taking effects from the environment into account. Taking the environment into account, which arises by including the whole measurement setup before and after the cavity (Fig.~\ref{figS:full_setup}), Equ.~\ref{equS:S21_ideal} becomes~\cite{Sprobst_efficient_2015}:
\begin{equation}
     S_{21}^{\text{full}} (\omega) = (a e^{i \alpha} e^{- i \omega \tau}) S_{21} (\omega)
     \label{equS:S21_full}
\end{equation}
Here $a$ and $\alpha$ are an additional attenuation and phase shift, independent of frequency and $\tau$ is the electrical phase delay of the microwave signal over the measurement setup.

\begin{figure*}[t]
    \centering
    \includegraphics[width = 14cm]{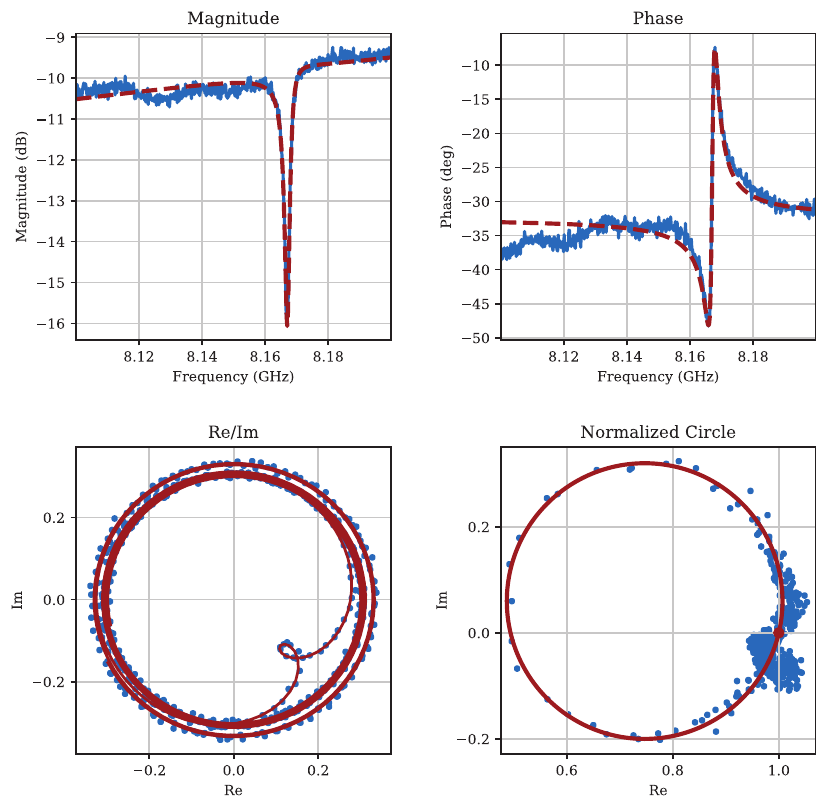}
    \caption{Circle fit to the U-shaped microwave cavity measured in transmission. Blue is the measurement data and red the fit with Equ.~\ref{equS:S21_full}. Top left: Direct magnitude $|S_{21}|$ of the measurement data and fit. Top right: Measured phase of $S_{21}$ with the electrical delay subtracted. Bottom left: Direct measurement data in the complex plane and fit of the full model. Bottom right: Data and fit in the complex plane after subtraction of the full environment.}
    \label{figS:full_circle_fit}
\end{figure*}
In Fig.~\ref{figS:full_circle_fit} we show the full circle fit data of the microwave cavity. The results of the fitting routine are given in table~\ref{tabS:fit_res_cav}.

\begin{table}[ht]
    \caption{Parameters obtained by a circle fit to the cavity. The internal quality factor $Q_{int}$ is calculated from the fit parameters, as $Q_l^{-1} = Q_c^{-1} + Q_{\text{int}}^{-1}$.}
    \centering
    \begin{tabular}{c|c|c|c|c|c|c|c|c}
    Fit parameter & $a$ & $\alpha$ & $\tau$ & $\omega_c/ 2 \pi$ & $\phi_0$ &  $Q_l$ & $Q_c$ & $Q_{\text{int}}$ \\ \hline
    Value & 0.32  & -0.56 rad & \SI{69.8}{ns} & \SI{8.1672}{GHz} & \SI{0.23}{rad} & 2913 & 5758  & 5898 
    \end{tabular}
    
    \label{tabS:fit_res_cav}
\end{table}
The measurement is performed at \SI{100}{mK} and at the highest frequency of the flux map. At this frequency the cavity is most insensitive to flux, hence we do not expect any influence of the cantilever on the parameters extracted.

\section{Details on the measurement protocol and data analysis}
\subsection*{Common procedure for all data}
Here describe how we measure and analyse the data. Several steps are common for all measurements. More specific routines required for individual measurements are described afterwards. For the measurements of the back action, there are slight modifications to the common steps, pointed out in the dedicated sections.

For all the measurements of the mechanical resonator we switched off the Pulse Tube Cooler of the cryostat, which gives us a measurement time window of around 10 minutes every 40 minutes. We measure the mechanical resonator with the spectrum analyser using a instrument bandwidth of \SI{0.1}{Hz}. We use a span of \SI{800}{Hz} centred around the mechanical frequency and set the number of points according to the bandwidth to 8001. It takes approximately \SI{10}{s} to take a single trace and we take 40 consecutive traces in one measurement window. We then average 4 traces on top of each other and fit them with the model for a damped harmonic oscillator Eq.~\ref{equS:PSD_HO}. We extract the coupling, using the calibration tone. The calibration peak is within the measurement span of the spectrum and therefore measured simultaneously. For all measurements, except the back action measurements, we take a VNA scan before and after the spectrum analyser measurements. This is done to ensure that the cavity did not move in frequency due to flux noise during the data acquisition. For the back action measurements, we take a low power VNA sweep simultaneously with each spectrum to determine the exact cavity frequency.

\subsection*{Details on measuring the increasing phonon number with temperature}
\begin{figure*}[h]
    \centering
    \includegraphics[width = 17.2cm]{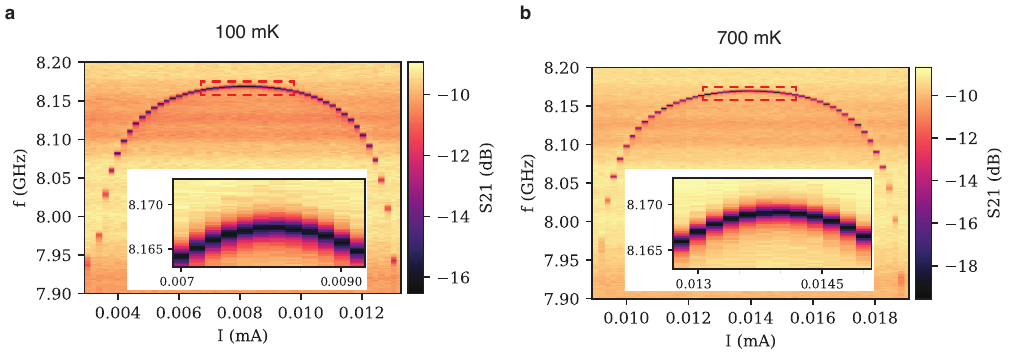}
    \caption{Flux maps at (a) \SI{100}{mK} and (b) \SI{700}{mK} over one flux period, the flux point is controlled via the coils around the waveguide. The insets show the magnification of the top of the flux map. The flux maps were taken with the Pulse Tube Cooler switched on.}
    \label{figS:fluxMap}
\end{figure*}

\begin{figure*}[t]
    \centering
    \includegraphics[width = 8.6cm]{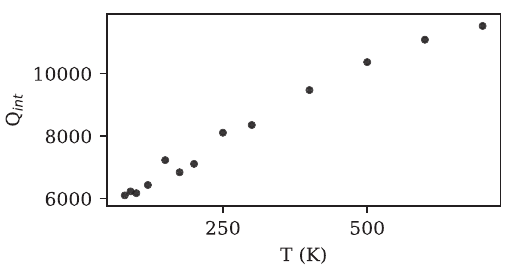}
    \caption{Dependence of the  microwave cavity internal quality factor on the cryostat temperature.}
    \label{figS:Qi_temp}
\end{figure*}

The temperature ramp in Figure 2a of the main body demonstrates that the mechanical system is thermalized, which allows us to evaluate the phonon number for a given temperature. This is necessary since in the experiment we only have access to $g_0 \sqrt{n}$, with $n$ being the phonon number: the knowledge of the phonon number is required to extract $g_{0}$. To trust the extracted values, the coupling $g_0$ has to be the same for measurements at different temperatures. In Figure 2b of the main body, we show that the coupling depends on the flux bias point. Hence we have to work at the same flux bias point at all temperatures. We observe that the maximum of the flux maps increases with temperature. To compensate for this, we perform the measurements \SI{15.6}{MHz} below the maximum frequency at each temperature. This ensures that $g_0$ is the same for all measured temperatures. The flux maps for \SI{100}{mK} and \SI{700}{mK} are shown in Fig.~\ref{figS:fluxMap}, where the shift with temperature is clearly visible comparing the insets. Furthermore we see that the linewidth reduces for higher temperatures. Something which has been observed and is usually explained with the saturation of lossy two level defects by temperature~\cite{Szoepfl_characterization_2017}. This leads to an increase of the internal quality factor with cryostat temperature, shown in Fig.~\ref{figS:Qi_temp}.

At \SI{100}{mK} we extract a maximum cavity frequency of \SI{8.1676}{GHz} by performing a circle fit, which gives a measurement frequency of \SI{8.152}{GHz}. At \SI{700}{mK} we extract a highest frequency of \SI{8.1694}{GHz}.

\begin{figure*}[t]
    \centering
    \includegraphics[width = 8.5cm]{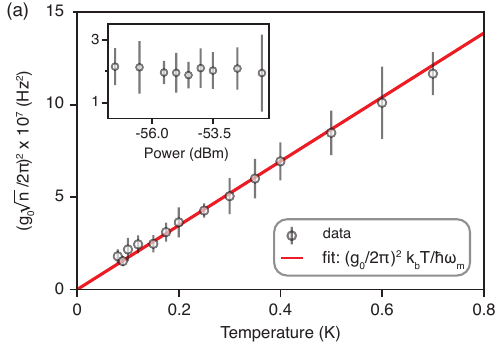}
    \caption{Measurement of $g_0 \sqrt{n}$ with increasing cryostat temperature at a fixed sensitivity with a fridge input power of \SI{-54.5}{dBm}. The inset shows the independence of the value $g_0 \sqrt{n}$ when changing the input power from \SI{-57.5}{dBm} to \SI{-51.5}{dBm} at \SI{100}{mK}. This Figure without the inset is also shown in Fig.~2(a) of the main manuscript.}
    \label{figS:tempRampg0}
\end{figure*}

For the temperature ramp, we set the microwave probe tone on resonance with the cavity. We ensure that the measurement power is low enough to avoid any back action on the mechanical system. We verify this by showing that $g_0 \sqrt{n}$ remains constant for varying the fridge input power from \SI{-57.5}{dBm} to \SI{-51.5}{dBm} at \SI{100}{mK}, while the measurements at different temperature are all done at \SI{-54.5}{dBm}. This is shown in as in inset to the full temperature ramp in Figure~\ref{figS:tempRampg0}.
The comparably low coupling (low flux sensitivity) limits the effect flux noise. We perform the measurements for each temperature over one Pulse Tube off period, which gives us 10 values after averaging in groups of 4 the spectrum traces. These 10 values are used to build up statistics. The error shown in the plot of the main manuscript is the standard deviation of those 10 values. We took the measurements for the temperature ramp over several days in a non-ordered fashion.

\subsection*{Details on measuring the dependence of $g_0$ on the flux bias point}
Here we measure the increase of the bare coupling strength, $g_0$, by changing the flux sensitivity of the microwave cavity. The back action on the mechanical cantilever increases linearly with the coupling and also depends on the number of circulating photons, $g_0 \sqrt{n_{photon}}$~\cite{Saspelmeyer_cavity_2014}. As a result, to faithfully extract the bare coupling $g_0$ we have to work in a regime without back action, as we only have access to $g_0 \sqrt{n}$ (with $n$ being the number of phonons) in the measurement and back action would change the phonon number. To ensure no back action, we lower $n_{photon}$ as we increase the flux sensitivity to keep $g_0 \sqrt{n_{photon}}$ approximately constant. We approximate the change of $g_0$ by the slope of the flux map and reduce the input power accordingly. In table~\ref{tabS:dBm_g0_fluxMap} we give the input powers for all measurements. For the more flux sensitive points, we took several measurements with slightly different power due to the poor signal to noise ratio. An important measure for back action is the linewidth of the mechanical resonator, which changes in case of back action. Hence we made sure that the linewidth remains the same for all the measurements (Fig.~\ref{figS:freqLWChange_g0_fluxMap}(a)).

\begin{table}[ht]
    \caption{Cryostat input powers for the measurement of the adjustable flux point. For some frequencies we performed several measurements with different power. In this case we give the range of the used powers.}
    \centering
    \begin{tabular}{c|c}
    Cavity frequency ($\omega_c/2\pi$) & Fridge input power \\ \hline
    \SI{8.15}{GHz} & \SI{-54.5}{dBm} \\ \hline
    \SI{8.12}{GHz} & \SI{-63.3}{dBm} \\ \hline
    \SI{8.102}{GHz} & \SI{-66.5}{dBm} \\ \hline
    \SI{8.046}{GHz} & \SI{-72.5}{dBm} \\ \hline
    \SI{7.996}{GHz} & \SI{-78.5}{dBm} \\ \hline
    \SI{7.936}{GHz} & \SI{-82.5}{dBm} \\ \hline
    \SI{7.928}{GHz} & \SI{-84.5}{dBm} to \SI{-81.5}{dBm} \\ \hline
    \SI{7.874}{GHz} & \SI{-86.5}{dBm} to \SI{-81.5}{dBm} \\ \hline
    \SI{7.858}{GHz} & \SI{-85.5}{dBm} to \SI{-81.5}{dBm}
    \end{tabular}
    
    \label{tabS:dBm_g0_fluxMap}
\end{table}

For this measurement, we set the probe tone in resonance with the microwave cavity and again average four data traces before fitting. Due to the low signal to noise ratio and high flux (noise) sensitivity, especially for the data with high coupling, it is necessary to apply rejection criteria. We apply those criteria to all data, and only consider data, which passes all criteria. We require, that the highest data point must not be more than \SI{4}{dB} from the maximum of the fit. Furthermore, the maximum of the fit must be at least \SI{4}{dB} above the noise floor. Lastly, we ignore a complete data set if more than 25\% fail the two criteria. Hence we enforce at least 7 averaged data traces for each measured coupling, which we use to build up statistics (the error shown in the plot of the main manuscript is the standard deviation of those points). These criteria ensure that we can trust the data points we extract from the fit. Small modifications of the criteria show the same (qualitative) results. The main limitation for the measurements at high $g_0$ is the low signal to noise. Additionally, flux noise might slightly detune the cavity frequency, which reduces the signal further. When measuring on resonance with the cavity, the calibration tone is strong enough, despite the low input power.

\subsection*{Details on the backaction measurement in the weak coupling regime}
For this measurement we measure the mechanical cantilever with a probe tone sweeping the cavity resonance. We do this at higher powers to get back action. The coupling is at a similar value as for the temperature ramp, which makes flux noise during the measurement not significant. For each measurement, to ensure we are operating at the same flux bias (compensating drifts over the full day) we tune the cavity frequency using the VNA before each measurement while the probe tone switched on (always at the same frequency). For the measurements itself we detune the probe tone to the intended frequency. We average at least 4 traces on top of each other before fitting. We apply the same goodness of fit and data criteria as for the measurements on the adjustable flux point, which are that the maximum of the data and the maximum of the fit are not allowed to be more than \SI{4}{dB} apart and the maximum of the fit has to be at least \SI{4}{dB} above the noise floor. We also require that for each detuning at least one averaged data trace consisting of four consecutive traces pass those criteria. We additionally check the average height of the calibration tone, which has to be above \SI{-130}{dBm}. Especially in regions far off resonance, the transduction of the microwave cavity is low, which limits the height of the calibration tone. In addition there can be spurious effects on the calibration peak. This can be due to slightly different electrical lengths of the cable through the experiment to the RF port and of the cable to the LO port of the mixer. We also observed leakage in the mixer itself, which leads to a spurious calibration peak as well. Those effects motivate to put a constraint on the calibration peak.
In contrast to all the other measurements, we used the spectrum analyser with a \SI{0.2}{Hz} bandwidth. This doubles the amount of data we can take within one Pulse Tube off measurement window. The error shown in the main body is the propagated fit error. Using the standard deviation instead, as for the measurements without back action, gives qualitative very similar errors.

\subsection*{Details on the backaction measurement in the large coupling regime}
\begin{figure*}[t]
    \centering
    \includegraphics[width = 8.6cm]{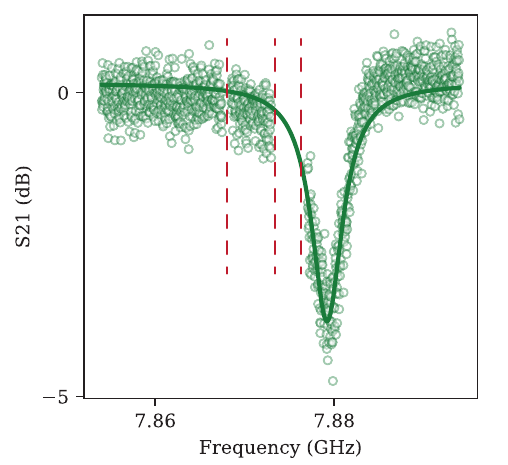}
    \caption{Fit of the cavity resonance with a Lorentzian. Indicated by the dashed lines we cut out a \SI{2}{MHz} window around the pump and a small window \SI{6}{MHz} below the pump, where the VNA shows a spurious pump component. The circles are the measured data, the line is the fit.}
    \label{figS:cavFit_high_g0}
\end{figure*}

\begin{figure*}[t]
    \centering
    \includegraphics[width = 8.6cm]{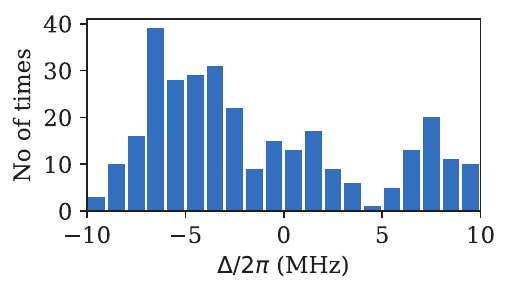}
    \caption{Grouping of the data traces in the case of high coupling. Each bin gives one data point for the back action measurement. To fit the data, at least four traces are averaged on top of each other. Here we plot a histogram of the bin sizes against the detuning to the cavity resonance.}
    \label{figS:histo_high_g0}
\end{figure*}

For the back action measurement with large coupling, where the system is highly flux sensitive, we suffer from flux drifts, which do not allow for a stable operation over the full 10 minutes measurement time. A first modification is that we only take 10 spectrum analyser traces instead of 40. In addition we acquire a VNA trace in parallel to the spectrum. To avoid any impact of the VNA measurement, we measure with around \SI{10}{dB} less than the probe tone. We fit this VNA traces with a Lorentzian to extract the resonance frequency of the microwave cavity (Fig.~\ref{figS:cavFit_high_g0}) for each data trace. Doing this fit, we cut out a \SI{2}{MHz} window around the pump, as well as a small window \SI{6}{MHz} below the pump frequency as the VNA shows a spurious pump component there. As the signal to noise ratio is low, especially in the regions with larger detunings, additional averaging is required. We do this by grouping the data with similar detuning in bins and again averaging at least four traces on top of each other. We decided for bin sizes of \SI{1}{MHz}. The size of each bin against the detuning is plotted in Fig.~\ref{figS:histo_high_g0}.
To obtain the frequency detuning of each bin, we take the average of the detunings in each bin, known from the Lorentzian fit. Changing the sizes of the bins does not give any qualitative difference on the data. The standard deviation of the detunings to the mean value is used as a x-error for each bin.
Similar to before, we perform several checks, which we apply to all the data. If the data fails one of the checks, we reject it in the analysis. We check the calibration tone, the goodness of the fit and require a bin to contain at least four data traces.
The calibration tone suffers from similar issues than in the weak coupling measurement, which are even more pronounced due to the weaker probe power. To test if there is a spurious calibration peak, we detune the microwave cavity after each measurement to have ideally no calibration peak, increase the development of the frequency modulation from \SI{200}{kHz} to \SI{500}{kHz} and measure the spurious calibration peak. If it is above \SI{-130}{dBm} we disregard the data set all together. We also re-calibrate if necessary before each measurement using the variable phase shifter. For each bin we also require that the mean of the calibration is above \SI{-130}{dBm}, where we used a development of \SI{200}{kHz}. Similar to before, we test the goodness of the fit by limiting the distance between the highest data point and maximum of the fit to \SI{4}{dB}. Furthermore, we require that the maximum of the fit has to be at least \SI{4}{dB} above noise floor. Only if a data point (consisting of at least four traces) passes all of these tests, we consider it in the analysis. As in the weak coupling we use the propagated fit error as the error for the y-axis.

\section{Dependence of the mechanical frequency and the mechanical linewidth on temperature}
\begin{figure*}[h]
    \centering
    \includegraphics[width = 15.3cm]{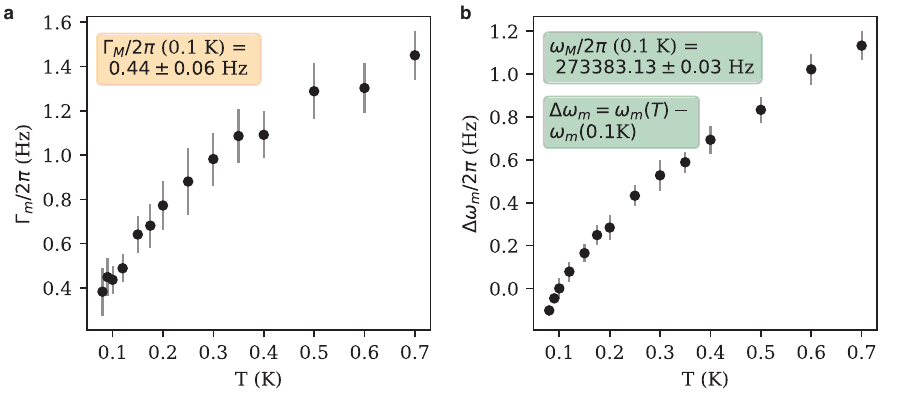}
    \caption{(a) Change of the mechanical linewidth with temperature. (b) Change of the mechanical frequency with temperature, relative to the frequency measured at \SI{100}{mK}. We extract this data by fitting the power spectrum with the model for a damped harmonic oscillator~\cite{Szhou_slowing_2013}. The error is the standard deviation of 10 data points, each obtained by fitting four averaged traces.}
    \label{figS:freqLWChange_tempRamp}
\end{figure*}

Next to an increase in phonon number, when increasing the temperature of the cryostat, we also observe an increase in linewidth and shift in frequency of the mechanical cantilever, Fig.~\ref{figS:freqLWChange_tempRamp}. In the measured temperature range from \SI{80}{mK} to \SI{700}{mK} we measure a more than three fold increase of the mechanical linewidth. A similar trend has been observed in other mechanical systems~\cite{Syuan_silicon_2015}, explained with a change of material properties. We also measure a change of mechanical frequency with temperature, which is likely also related to changing material properties.

\section{Dependence of the cavity and mechanical parameters on the flux bias point}
\subsection*{Mechanical linewidth and frequency in dependence of the coupling}
\begin{figure*}[t]
    \centering
    \includegraphics[width = 15.3cm]{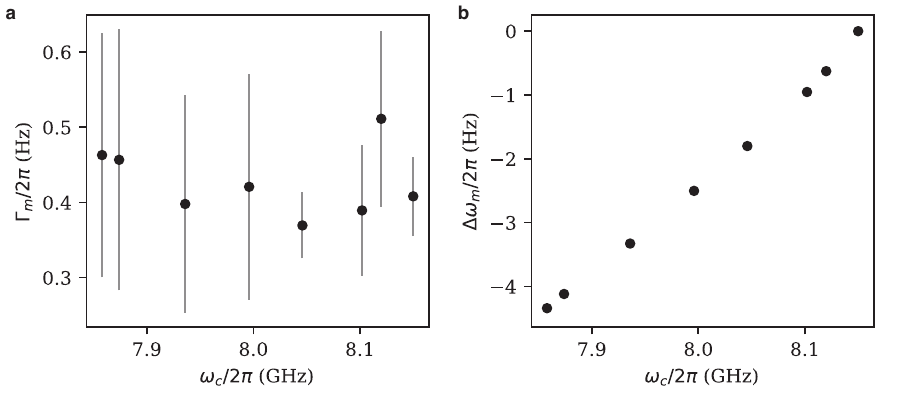}
    \caption{(a) Linewidth of the mechanical cantilever measured at different flux bias points. The constant linewidth proves that we do not induce back action, which ensures the mechanical resonator remains thermal, a required assumption to extract the value of $g_0$. (b) Change of the mechanical frequency with coupling strength. See main text for explanation. The error in both panels is the standard deviation to the mean value of multiple measurements. In (b) it is too small to be visible.}
    \label{figS:freqLWChange_g0_fluxMap}
\end{figure*}

Here we show the mechanical linewidth and frequency shift with the coupling, $g_0$, which is set by the flux bias point of the microwave cavity. A change in the linewidth with the coupling strength would imply that we induce back action on the mechanical system by measuring with too high power. In Fig.~\ref{figS:freqLWChange_g0_fluxMap}(a) we see that the linewidth remains constant at the value we expect for no back action. This proves that we do not induce back action and we can trust the coupling value we extract since it assumes a thermalized mechanical mode. 
In Fig.~\ref{figS:freqLWChange_g0_fluxMap}(b) we show the change of the mechanical frequency with the coupling strength. Despite the cantilever being a mechanical resonator, it can be modelled as an LC resonator which results in a system of two inductively coupled LC resonators. In case one inductance changes - which is the case when we tune the frequency of the microwave cavity - it leads to a change of the effective inductance of the other resonator. This leads to a frequency change and explains why the frequency of the mechanical resonator depends on the frequency of the microwave cavity.

\subsection*{Cavity linewidth and cooperativity in dependence of the coupling}
Here we show how the cavity linewidth changes with the flux bias point (which sets the coupling strength). In combination with the change of coupling (see main manuscript), this allows to extract the cooperativity for each coupling point.

\begin{figure*}[t]
    \centering
    \includegraphics[width = 15.3cm]{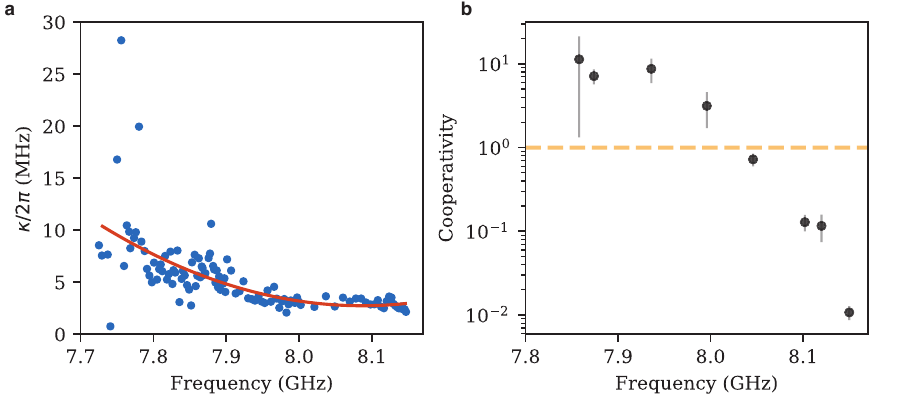}
    \caption{(a) Linewidth of the microwave cavity against its frequency which is adjusted by the flux bias point. We fit this using a polynomial fit (red line). The linewidth increases due to increasing sensitivity to flux noise. (b) Single-photon cooperativity against cavity frequency (flux bias point). We extract $g_0$ from the measurement shown in the main paper, which allows to calculate the cooperativity (solid line) in combination with the information shown in (a). The data points are the measured cooperativity using the extracted coupling and microwave linewidth at this point and the mechanical linewidth. The dashed line shows where the single-photon cooperativity exceeds unity.}
    \label{figS:g0_lwmw_c}
\end{figure*}

We extract the linewidth of the cavity in dependence of the flux bias point by circle fitting the individual measurements of the flux map. To record this flux map, we switched off the Pulse Tube Cooler, to avoid other excitation of the cantilever than from its thermal environment. We show the measured linewidth in Fig.~\ref{figS:g0_lwmw_c}. The linewidth for the most flux sensitive points increases from around \SI{3}{MHz} at the most flux insensitive point to around \SI{10}{MHz}. From the cantilever we expect a broadening of the linewidth due to its impact given by $g_0/2 \pi \sqrt{n} \simeq \SI{0.25}{MHz}$ at the highest coupling with $g_0/2 \pi \simeq \SI{3}{kHz}$. Hence the broadening of the linewidth is substantially bigger than expected from the cantilever and we attribute this to the increased sensitivity to flux noise. This also reduces the cooperativity, which directly depends on the linewidth.

The single-photon cooperativity against the flux bias point (microwave cavity frequency) is shown in Fig.~\ref{figS:g0_lwmw_c}(b). The cooperativity is given by $C_0 = 4 g_0^2 / (\kappa \Gamma_m)$, thus it is directly influenced by the increasing cavity linewidth $\kappa$ with the increasing flux sensitivity. For the data points shown in the plot, we extract the cooperativity using the direct measurement of $g_0$ as well as $\kappa$ from a direct VNA measurement. For the solid line, we use the fit of $g_0$ against the flux point (see main manuscript) as well as the fitted cavity linewidth shown in panel (a). The dashed line indicates where we reach single-photon cooperativity exceeding unity.

\section{Details on fitting the measurement data on back action}

\subsection{Equations and additional details about the fitting routine}

For the cooling traces presented in the main body we fitted the phonon number using the conventional theory for cavity cooling assuming a linear cavity~\cite{Ssafavi-naeini_laser_2013}. Concretely, the optical springing and damping terms are given:
\begin{equation}
\begin{split}
    &\delta \omega_m = \left| G\right|^2 \text{Im} \left[ \frac{1}{\text{i}\left( \Delta - \omega_m \right) + \kappa/2} -  \frac{1}{-\text{i}\left( \Delta + \omega_m \right) + \kappa/2} \right], \\
    &\gamma_{OM} = 2\left| G\right|^2 \text{Re} \left[ \frac{1}{\text{i}\left( \Delta - \omega_m \right) + \kappa/2} -  \frac{1}{-\text{i}\left( \Delta + \omega_m \right) + \kappa/2} \right],
\end{split}
\label{eqSM:spring}
\end{equation}
where $G = g_0 \sqrt{n_{ph}\left(\omega\right)}$ is the photon enhanced coupling for a pump frequency $\omega_p$, $\kappa$ the cavity linewidth, and $\Delta = \omega_c - \omega_p - \omega_{off}$ the detuning from cavity frequency $\omega_c$. We included an additional frequency offset, $\omega_{off}$, to accommodate for the impedance mismatch of the cavity to the waveguide. The number of phonons in the mechanics is then given by:
\begin{equation}
    n_f\left( \omega \right) = \frac{\gamma_i n_b}{\gamma} + \frac{\left| G \right|^2 \kappa}{\gamma} \frac{1}{\left(\Delta - \omega \right)^2 + \kappa^2/4},
    \label{eqSM:phonon}
\end{equation}
where $\gamma_i$ is the intrinsic mechanical linewidth, $\gamma = \gamma_i + \gamma_{OM}$ the total mechanical linewidth (which would be $\Gamma_M$ in our experimental data), and $n_b$ the average thermal bath occupancy. The second term is the residual heating from non-resonant scattering of photons. In the weak-coupling regime ($\kappa \gg \gamma$) this quantum limit on backaction cooling is $n_{qbl} = (\kappa/4\omega_m)^2$.

From our experimental data, we have direct access to the photon enhanced coupling $g_0\sqrt{n}$ which we fit using:
\begin{equation}
    \left. g_0^2 n\left( \omega \right) \right|_{exp} = \frac{g_0^2}{\gamma_i + 2 g_0^2 n_{ph} \left(\omega\right) \text{Re} \left[ \frac{1}{\text{i}\left( \Delta - \omega_m \right) + \kappa/2} -  \frac{1}{-\text{i}\left( \Delta + \omega_m \right) + \kappa/2} \right]} \left[ \gamma_i n_b + \frac{\kappa g_0^2 n_{ph}\left(\omega\right)}{\left(\Delta - \omega \right)^2 + \kappa^2/4} \right],
\end{equation}
where the circulating photon number $n_{ph}\left(\omega\right)$ is $n_0/(4\Delta^2/\kappa^2+1) $, reaching a maximum circulating photon number on resonance of $n_0$. In Figure~3 and 4, this function was used with $g_0$ and $n_0$ as fit parameters while the cavity linewidth $\kappa$, the intrinsic mechanical frequency $\omega_m$ and linewidth $\gamma_i$ were fixed using prior measurements at low power (no backaction, e.g. in Figure~2). This approach allows us to obtain an independent value of the coupling strength $g_0$ as well as the circulating photon number in the cavity on resonance $n_0$. Using theses parameter, we also show the theoretical prediction for the mechanical linewidth and frequency in Figures~3 and 4 of the main body. We also note that this theoretical formalism is correct for $\gamma>0$, i.e. away from the instability region.

\subsection{Effect of the flux noise on the mechanical linewidth}
In Figure~4 of the main body, we show the measurement data on back action in case of large coupling and compare it to theory. The theoretical fit obtained from the phonon number and the prediction for the mechanical frequency are in good agreement with the measurement results. However the theoretical prediction for mechanical linewidth is significantly smaller than the measurement, in particular in the region of strong cooling. This can be explained with flux noise and how this interacts with the mechanical frequency shift and increases the linewidth. 

In Figure~\ref{figS:g0_lwmw_c}(a) we show that the linewidth of the cavity increases with the flux bias point due to flux noise. We attribute this to excess current noise in the lines feeding the coils used to bias the cavity and effectively change the cavity frequency on timescales much shorter than the measurement. In the context of the back action measurement, this noise can be interpreted as very rapid changes of the exact detuning of the cavity. Due to the optical spring effect on the mechanics, the cavity flux noise will in turn induce noise on the mechanical frequency that will further broaden the measured power spectrum.

\begin{figure*}[t]
    \centering
    \includegraphics[width = 6cm]{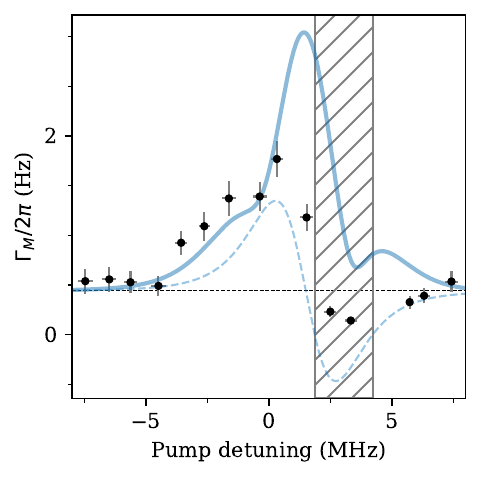}
    \caption{Mechanical linewidth when taking flux noise into account. The frequency of the cavity changes rapidly in time due to flux noise, which leads to an increase in the measured mechanical linewidth due to the optical spring effect. The dashed line is the theoretical prediction as plotted in the main body, with this effect not taken into account. The solid line is the theoretical prediction taking the frequency fluctuations of the cavity into account using a simple toy model.}
    \label{figS:gam_detail_fluxNoise}
\end{figure*}

With a very simple model we can estimate the impact the flux noise has on the measured mechanical linewidth. Indeed, for a given regime (coupling $g_0$ and circulating photon number on resonance $n_0$) we can calculate how much the mechanical frequency will change for a given detuning change by using the derivative of the optical spring effect (eq.~\ref{eqSM:spring}). Concretely, based on the parameters from the data presented in Figure~4 of the main body, we generated a normally distributed noise around each of the detuning values. Using the theory for the optical spring effect we then calculate how this noise changes the mechanical frequencies during the measurement. The standard deviation from the resulting distribution is then finally used along with the theoretical optical damping to predict the overall mechanical linewidth (assuming the noise distribution from the mechanical frequency and the optical damping can be summed as uncorrelated normal distributions). This simple toy-model produces an increased mechanical linewidth which is in good qualitative agreement with the linewidth observed in the experiment, as shown in Figure~\ref{figS:gam_detail_fluxNoise}. We note that in the vicinity of the instability region (hashed area in Figure~\ref{figS:gam_detail_fluxNoise}) this toy-model cannot be trusted since the theory for the mechanical back action is not valid. In order to further improve upon such toy-model, it is necessary to accurately model the source and spectral characteristics of the flux noise. This will indeed directly impact how the mechanical frequency is shifted as a function of time and the resulting increase of the mechanical linewidth.

\section{Estimation of the maximum photon number in the microwave cavity}
The strength of the back action on the mechanical cantilever, and hence the lowest phonon occupation we can cool to, depends on the number of photons in the microwave cavity: the more photons the stronger the cooling (for a given coupling strength). This is only valid until we reach the back action limit, which is given by $\left< n ^{\text{min}} \right> = (\kappa / 4 \omega_m)^2 \simeq 6.5$~\cite{Smarquardt_quantum_2007}. However this assumes an arbitrarily high photon number in the microwave cavity, which is not possible in practice. Since our approach relies on a SQUID to mediate the optomechanical coupling, the cavity is intrinsically non-linear and can be described in the framework of the Duffing model~\cite{Snation_quantum_2008}. Such a Duffing resonator becomes bistable beyond a certain input pump strength. As we want to operate below this bistability, this imposes a maximum input power, thus photon number in the cavity. The maximum number of photons is directly related to the strength of the non-linearity and the cavity linewidth as follows~\cite{Sagrawal_optical_1979}:
\begin{equation}
    n_{\text{ph}}^{\text{crit}} = \frac{\kappa}{\sqrt{3} K}.
    \label{equS:Duff_crit}
\end{equation}
Here $n_{\text{ph}}^{\text{crit}}$ is the critical photon number beyond which the cavity become bistable, $\kappa$ is the cavity linewidth and $K$ the Kerr shift per input photon, which is linear with input power.

\begin{figure*}[t]
    \centering
    \includegraphics[width = 15cm]{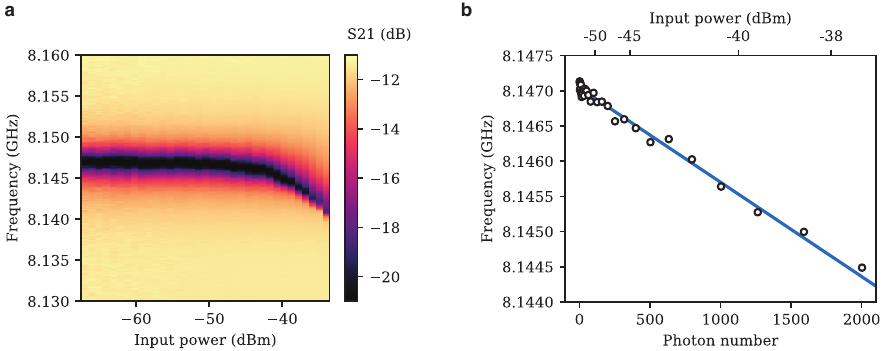}
    \caption{Shift of the cavity frequency with input power. (a) Change of the cavity frequency obtained by transmission measurements for increasing VNA input powers. (b) Extracted frequency for each input power using the circle fit routine (see section~4). The input power is converted to photon number, as the frequency shift has a linear dependence on input power. We fit that that change, to extract a Kerr constant $K = \SI{-1.3}{kHz}$ per photon.}
    \label{figS:kerr_shift}
\end{figure*}

In Figure~\ref{figS:kerr_shift} we plot the frequency shift against the input power. In (a) we plot a map of the response and in (b) the extracted frequency shift with photon number in the cavity. The photon number is evaluated using the extracted photon number from the back action measurement shown in Fig.~3 and the changing fridge input power. We fit a linear slope to it and extract a Kerr constant $K = \SI{-1.3}{kHz}$ per photon. Using this in combination with equation~\ref{equS:Duff_crit} we extract a critical photon number of $\simeq \SI{1200}{}$ before the cavity becomes bistable. We note that the data presented in Fig.~\ref{figS:kerr_shift} is from another cool down, but measured on the same sample. The coupling of the cavity to the waveguide was slightly different, which we carefully take into account for this estimation. This is also a conservative bound, as the measurement was taken at a flux point \SI{2.5}{MHz} lower than the cooling curve and the Kerr increases with the flux sensitivity. Thus we are slightly under-estimating the maximum photon number. Also, we want to emphasise, that this estimation is valid for the low coupling point of $g_0 / 2 \pi \simeq \SI{57}{Hz}$. Measuring at a more flux sensitive point, increases the back action for a given photon number due to an increases in coupling, but at the same time the Kerr shift increases, leading to a lower maximum photon number before the cavity becomes bistable.

\end{CJK*}


\begin{thebibliography}{42}%
\makeatletter
\providecommand \@ifxundefined [1]{%
 \@ifx{#1\undefined}
}%
\providecommand \@ifnum [1]{%
 \ifnum #1\expandafter \@firstoftwo
 \else \expandafter \@secondoftwo
 \fi
}%
\providecommand \@ifx [1]{%
 \ifx #1\expandafter \@firstoftwo
 \else \expandafter \@secondoftwo
 \fi
}%
\providecommand \natexlab [1]{#1}%
\providecommand \enquote  [1]{``#1''}%
\providecommand \bibnamefont  [1]{#1}%
\providecommand \bibfnamefont [1]{#1}%
\providecommand \citenamefont [1]{#1}%
\providecommand \href@noop [0]{\@secondoftwo}%
\providecommand \href [0]{\begingroup \@sanitize@url \@href}%
\providecommand \@href[1]{\@@startlink{#1}\@@href}%
\providecommand \@@href[1]{\endgroup#1\@@endlink}%
\providecommand \@sanitize@url [0]{\catcode `\\12\catcode `\$12\catcode
  `\&12\catcode `\#12\catcode `\^12\catcode `\_12\catcode `\%12\relax}%
\providecommand \@@startlink[1]{}%
\providecommand \@@endlink[0]{}%
\providecommand \url  [0]{\begingroup\@sanitize@url \@url }%
\providecommand \@url [1]{\endgroup\@href {#1}{\urlprefix }}%
\providecommand \urlprefix  [0]{URL }%
\providecommand \Eprint [0]{\href }%
\providecommand \doibase [0]{https://doi.org/}%
\providecommand \selectlanguage [0]{\@gobble}%
\providecommand \bibinfo  [0]{\@secondoftwo}%
\providecommand \bibfield  [0]{\@secondoftwo}%
\providecommand \translation [1]{[#1]}%
\providecommand \BibitemOpen [0]{}%
\providecommand \bibitemStop [0]{}%
\providecommand \bibitemNoStop [0]{.\EOS\space}%
\providecommand \EOS [0]{\spacefactor3000\relax}%
\providecommand \BibitemShut  [1]{\csname bibitem#1\endcsname}%
\let\auto@bib@innerbib\@empty
\bibitem [{\citenamefont {O{\textquoteright}Connell}\ \emph
  {et~al.}(2010)\citenamefont {O{\textquoteright}Connell}, \citenamefont
  {Hofheinz}, \citenamefont {Ansmann}, \citenamefont {Bialczak}, \citenamefont
  {Lenander}, \citenamefont {Lucero}, \citenamefont {Neeley}, \citenamefont
  {Sank}, \citenamefont {Wang}, \citenamefont {Weides}, \citenamefont {Wenner},
  \citenamefont {Martinis},\ and\ \citenamefont
  {Cleland}}]{oconnell_quantum_2010}%
  \BibitemOpen
  \bibfield  {author} {\bibinfo {author} {\bibfnamefont {A.~D.}\ \bibnamefont
  {O{\textquoteright}Connell}}, \bibinfo {author} {\bibfnamefont
  {M.}~\bibnamefont {Hofheinz}}, \bibinfo {author} {\bibfnamefont
  {M.}~\bibnamefont {Ansmann}}, \bibinfo {author} {\bibfnamefont {R.~C.}\
  \bibnamefont {Bialczak}}, \bibinfo {author} {\bibfnamefont {M.}~\bibnamefont
  {Lenander}}, \bibinfo {author} {\bibfnamefont {E.}~\bibnamefont {Lucero}},
  \bibinfo {author} {\bibfnamefont {M.}~\bibnamefont {Neeley}}, \bibinfo
  {author} {\bibfnamefont {D.}~\bibnamefont {Sank}}, \bibinfo {author}
  {\bibfnamefont {H.}~\bibnamefont {Wang}}, \bibinfo {author} {\bibfnamefont
  {M.}~\bibnamefont {Weides}}, \bibinfo {author} {\bibfnamefont
  {J.}~\bibnamefont {Wenner}}, \bibinfo {author} {\bibfnamefont {J.~M.}\
  \bibnamefont {Martinis}},\ and\ \bibinfo {author} {\bibfnamefont {A.~N.}\
  \bibnamefont {Cleland}},\ }``Quantum ground state and single-phonon control
  of a mechanical resonator'',\ \href {https://doi.org/10.1038/nature08967}
  {\bibfield  {journal} {\bibinfo  {journal} {Nature}\ }\textbf {\bibinfo
  {volume} {464}},\ \bibinfo {pages} {697} (\bibinfo {year}
  {2010})}\BibitemShut {NoStop}%
\bibitem [{\citenamefont {Teufel}\ \emph {et~al.}(2011)\citenamefont {Teufel},
  \citenamefont {Donner}, \citenamefont {Li}, \citenamefont {Harlow},
  \citenamefont {Allman}, \citenamefont {Cicak}, \citenamefont {Sirois},
  \citenamefont {Whittaker}, \citenamefont {Lehnert},\ and\ \citenamefont
  {Simmonds}}]{teufel_sideband_2011}%
  \BibitemOpen
  \bibfield  {author} {\bibinfo {author} {\bibfnamefont {J.~D.}\ \bibnamefont
  {Teufel}}, \bibinfo {author} {\bibfnamefont {T.}~\bibnamefont {Donner}},
  \bibinfo {author} {\bibfnamefont {D.}~\bibnamefont {Li}}, \bibinfo {author}
  {\bibfnamefont {J.~W.}\ \bibnamefont {Harlow}}, \bibinfo {author}
  {\bibfnamefont {M.~S.}\ \bibnamefont {Allman}}, \bibinfo {author}
  {\bibfnamefont {K.}~\bibnamefont {Cicak}}, \bibinfo {author} {\bibfnamefont
  {A.~J.}\ \bibnamefont {Sirois}}, \bibinfo {author} {\bibfnamefont {J.~D.}\
  \bibnamefont {Whittaker}}, \bibinfo {author} {\bibfnamefont {K.~W.}\
  \bibnamefont {Lehnert}},\ and\ \bibinfo {author} {\bibfnamefont {R.~W.}\
  \bibnamefont {Simmonds}},\ }``Sideband cooling of micromechanical motion to
  the quantum ground state'',\ \href {https://doi.org/10.1038/nature10261}
  {\bibfield  {journal} {\bibinfo  {journal} {Nature}\ }\textbf {\bibinfo
  {volume} {475}},\ \bibinfo {pages} {359} (\bibinfo {year}
  {2011})}\BibitemShut {NoStop}%
\bibitem [{\citenamefont {Chan}\ \emph {et~al.}(2011)\citenamefont {Chan},
  \citenamefont {Alegre}, \citenamefont {Safavi-Naeini}, \citenamefont {Hill},
  \citenamefont {Krause}, \citenamefont {Gr{\"o}blacher}, \citenamefont
  {Aspelmeyer},\ and\ \citenamefont {Painter}}]{chan_laser_2011}%
  \BibitemOpen
  \bibfield  {author} {\bibinfo {author} {\bibfnamefont {J.}~\bibnamefont
  {Chan}}, \bibinfo {author} {\bibfnamefont {T.~P.~M.}\ \bibnamefont {Alegre}},
  \bibinfo {author} {\bibfnamefont {A.~H.}\ \bibnamefont {Safavi-Naeini}},
  \bibinfo {author} {\bibfnamefont {J.~T.}\ \bibnamefont {Hill}}, \bibinfo
  {author} {\bibfnamefont {A.}~\bibnamefont {Krause}}, \bibinfo {author}
  {\bibfnamefont {S.}~\bibnamefont {Gr{\"o}blacher}}, \bibinfo {author}
  {\bibfnamefont {M.}~\bibnamefont {Aspelmeyer}},\ and\ \bibinfo {author}
  {\bibfnamefont {O.}~\bibnamefont {Painter}},\ }``Laser cooling of a
  nanomechanical oscillator into its quantum ground state'',\ \href
  {https://doi.org/10.1038/nature10461} {\bibfield  {journal} {\bibinfo
  {journal} {Nature}\ }\textbf {\bibinfo {volume} {478}},\ \bibinfo {pages}
  {89} (\bibinfo {year} {2011})}\BibitemShut {NoStop}%
\bibitem [{\citenamefont {Teufel}\ \emph {et~al.}(2009)\citenamefont {Teufel},
  \citenamefont {Donner}, \citenamefont {Castellanos-Beltran}, \citenamefont
  {Harlow},\ and\ \citenamefont {Lehnert}}]{teufel_nanomechanical_2009}%
  \BibitemOpen
  \bibfield  {author} {\bibinfo {author} {\bibfnamefont {J.~D.}\ \bibnamefont
  {Teufel}}, \bibinfo {author} {\bibfnamefont {T.}~\bibnamefont {Donner}},
  \bibinfo {author} {\bibfnamefont {M.~A.}\ \bibnamefont
  {Castellanos-Beltran}}, \bibinfo {author} {\bibfnamefont {J.~W.}\
  \bibnamefont {Harlow}},\ and\ \bibinfo {author} {\bibfnamefont {K.~W.}\
  \bibnamefont {Lehnert}},\ }``Nanomechanical motion measured with an
  imprecision below that at the standard quantum limit'',\ \href
  {https://doi.org/10.1038/nnano.2009.343} {\bibfield  {journal} {\bibinfo
  {journal} {Nature Nanotechnology}\ }\textbf {\bibinfo {volume} {4}},\
  \bibinfo {pages} {820} (\bibinfo {year} {2009})}\BibitemShut {NoStop}%
\bibitem [{\citenamefont {Anetsberger}\ \emph {et~al.}(2010)\citenamefont
  {Anetsberger}, \citenamefont {Gavartin}, \citenamefont {Arcizet},
  \citenamefont {Unterreithmeier}, \citenamefont {Weig}, \citenamefont
  {Gorodetsky}, \citenamefont {Kotthaus},\ and\ \citenamefont
  {Kippenberg}}]{anetsberger_measuring_2010}%
  \BibitemOpen
  \bibfield  {author} {\bibinfo {author} {\bibfnamefont {G.}~\bibnamefont
  {Anetsberger}}, \bibinfo {author} {\bibfnamefont {E.}~\bibnamefont
  {Gavartin}}, \bibinfo {author} {\bibfnamefont {O.}~\bibnamefont {Arcizet}},
  \bibinfo {author} {\bibfnamefont {Q.~P.}\ \bibnamefont {Unterreithmeier}},
  \bibinfo {author} {\bibfnamefont {E.~M.}\ \bibnamefont {Weig}}, \bibinfo
  {author} {\bibfnamefont {M.~L.}\ \bibnamefont {Gorodetsky}}, \bibinfo
  {author} {\bibfnamefont {J.~P.}\ \bibnamefont {Kotthaus}},\ and\ \bibinfo
  {author} {\bibfnamefont {T.~J.}\ \bibnamefont {Kippenberg}},\ }``Measuring
  nanomechanical motion with an imprecision below the standard quantum
  limit'',\ \href {https://doi.org/10.1103/PhysRevA.82.061804} {\bibfield
  {journal} {\bibinfo  {journal} {Physical Review A}\ }\textbf {\bibinfo
  {volume} {82}},\ \bibinfo {pages} {061804} (\bibinfo {year}
  {2010})}\BibitemShut {NoStop}%
\bibitem [{\citenamefont {Wollman}\ \emph {et~al.}(2015)\citenamefont
  {Wollman}, \citenamefont {Lei}, \citenamefont {Weinstein}, \citenamefont
  {Suh}, \citenamefont {Kronwald}, \citenamefont {Marquardt}, \citenamefont
  {Clerk},\ and\ \citenamefont {Schwab}}]{wollman_quantum_2015}%
  \BibitemOpen
  \bibfield  {author} {\bibinfo {author} {\bibfnamefont {E.~E.}\ \bibnamefont
  {Wollman}}, \bibinfo {author} {\bibfnamefont {C.~U.}\ \bibnamefont {Lei}},
  \bibinfo {author} {\bibfnamefont {A.~J.}\ \bibnamefont {Weinstein}}, \bibinfo
  {author} {\bibfnamefont {J.}~\bibnamefont {Suh}}, \bibinfo {author}
  {\bibfnamefont {A.}~\bibnamefont {Kronwald}}, \bibinfo {author}
  {\bibfnamefont {F.}~\bibnamefont {Marquardt}}, \bibinfo {author}
  {\bibfnamefont {A.~A.}\ \bibnamefont {Clerk}},\ and\ \bibinfo {author}
  {\bibfnamefont {K.~C.}\ \bibnamefont {Schwab}},\ }``Quantum squeezing of
  motion in a mechanical resonator'',\ \href
  {https://doi.org/10.1126/science.aac5138} {\bibfield  {journal} {\bibinfo
  {journal} {Science}\ }\textbf {\bibinfo {volume} {349}},\ \bibinfo {pages}
  {952} (\bibinfo {year} {2015})}\BibitemShut {NoStop}%
\bibitem [{\citenamefont {Lecocq}\ \emph {et~al.}(2015)\citenamefont {Lecocq},
  \citenamefont {Clark}, \citenamefont {Simmonds}, \citenamefont {Aumentado},\
  and\ \citenamefont {Teufel}}]{lecocq_quantum_2015}%
  \BibitemOpen
  \bibfield  {author} {\bibinfo {author} {\bibfnamefont {F.}~\bibnamefont
  {Lecocq}}, \bibinfo {author} {\bibfnamefont {J.~B.}\ \bibnamefont {Clark}},
  \bibinfo {author} {\bibfnamefont {R.~W.}\ \bibnamefont {Simmonds}}, \bibinfo
  {author} {\bibfnamefont {J.}~\bibnamefont {Aumentado}},\ and\ \bibinfo
  {author} {\bibfnamefont {J.~D.}\ \bibnamefont {Teufel}},\ }``Quantum
  {Nondemolition} {Measurement} of a {Nonclassical} {State} of a {Massive}
  {Object}'',\ \bibfield  {journal} {\bibinfo  {journal} {Physical Review X}\
  }\textbf {\bibinfo {volume} {5}},\ \bibinfo {pages} {041037} \href
  {https://doi.org/10.1103/PhysRevX.5.041037}
  (\bibinfo {year} {2015})\BibitemShut {NoStop}%
\bibitem [{\citenamefont {Pirkkalainen}\ \emph
  {et~al.}(2015{\natexlab{a}})\citenamefont {Pirkkalainen}, \citenamefont
  {Damsk{\"a}gg}, \citenamefont {Brandt}, \citenamefont {Massel},\ and\
  \citenamefont {Sillanp{\"a}{\"a}}}]{pirkkalainen_squeezing_2015}%
  \BibitemOpen
  \bibfield  {author} {\bibinfo {author} {\bibfnamefont {J.-M.}\ \bibnamefont
  {Pirkkalainen}}, \bibinfo {author} {\bibfnamefont {E.}~\bibnamefont
  {Damsk{\"a}gg}}, \bibinfo {author} {\bibfnamefont {M.}~\bibnamefont
  {Brandt}}, \bibinfo {author} {\bibfnamefont {F.}~\bibnamefont {Massel}},\
  and\ \bibinfo {author} {\bibfnamefont {M.~A.}\ \bibnamefont
  {Sillanp{\"a}{\"a}}},\ }``Squeezing of {Quantum} {Noise} of {Motion} in a
  {Micromechanical} {Resonator}'',\ \href
  {https://doi.org/10.1103/PhysRevLett.115.243601} {\bibfield  {journal}
  {\bibinfo  {journal} {Physical Review Letters}\ }\textbf {\bibinfo {volume}
  {115}},\ \bibinfo {pages} {243601} (\bibinfo {year}
  {2015}{\natexlab{a}})}\BibitemShut {NoStop}%
\bibitem [{\citenamefont {Reed}\ \emph {et~al.}(2017)\citenamefont {Reed},
  \citenamefont {Mayer}, \citenamefont {Teufel}, \citenamefont {Burkhart},
  \citenamefont {Pfaff}, \citenamefont {Reagor}, \citenamefont {Sletten},
  \citenamefont {Ma}, \citenamefont {Schoelkopf}, \citenamefont {Knill},\ and\
  \citenamefont {Lehnert}}]{reed_faithful_2017}%
  \BibitemOpen
  \bibfield  {author} {\bibinfo {author} {\bibfnamefont {A.~P.}\ \bibnamefont
  {Reed}}, \bibinfo {author} {\bibfnamefont {K.~H.}\ \bibnamefont {Mayer}},
  \bibinfo {author} {\bibfnamefont {J.~D.}\ \bibnamefont {Teufel}}, \bibinfo
  {author} {\bibfnamefont {L.~D.}\ \bibnamefont {Burkhart}}, \bibinfo {author}
  {\bibfnamefont {W.}~\bibnamefont {Pfaff}}, \bibinfo {author} {\bibfnamefont
  {M.}~\bibnamefont {Reagor}}, \bibinfo {author} {\bibfnamefont
  {L.}~\bibnamefont {Sletten}}, \bibinfo {author} {\bibfnamefont
  {X.}~\bibnamefont {Ma}}, \bibinfo {author} {\bibfnamefont {R.~J.}\
  \bibnamefont {Schoelkopf}}, \bibinfo {author} {\bibfnamefont
  {E.}~\bibnamefont {Knill}},\ and\ \bibinfo {author} {\bibfnamefont {K.~W.}\
  \bibnamefont {Lehnert}},\ }``Faithful conversion of propagating quantum
  information to mechanical motion'',\ \href
  {https://doi.org/10.1038/nphys4251} {\bibfield  {journal} {\bibinfo
  {journal} {Nature Physics}\ }\textbf {\bibinfo {volume} {13}},\ \bibinfo
  {pages} {1163} (\bibinfo {year} {2017})}\BibitemShut {NoStop}%
\bibitem [{\citenamefont {Palomaki}\ \emph {et~al.}(2013)\citenamefont
  {Palomaki}, \citenamefont {Teufel}, \citenamefont {Simmonds},\ and\
  \citenamefont {Lehnert}}]{palomaki_entangling_2013}%
  \BibitemOpen
  \bibfield  {author} {\bibinfo {author} {\bibfnamefont {T.~A.}\ \bibnamefont
  {Palomaki}}, \bibinfo {author} {\bibfnamefont {J.~D.}\ \bibnamefont
  {Teufel}}, \bibinfo {author} {\bibfnamefont {R.~W.}\ \bibnamefont
  {Simmonds}},\ and\ \bibinfo {author} {\bibfnamefont {K.~W.}\ \bibnamefont
  {Lehnert}},\ }``Entangling {Mechanical} {Motion} with {Microwave}
  {Fields}'',\ \href {https://doi.org/10.1126/science.1244563} {\bibfield
  {journal} {\bibinfo  {journal} {Science}\ }\textbf {\bibinfo {volume}
  {342}},\ \bibinfo {pages} {710} (\bibinfo {year} {2013})}\BibitemShut
  {NoStop}%
\bibitem [{\citenamefont {Riedinger}\ \emph {et~al.}(2018)\citenamefont
  {Riedinger}, \citenamefont {Wallucks}, \citenamefont {Marinkovi{\'c}},
  \citenamefont {L{\"o}schnauer}, \citenamefont {Aspelmeyer}, \citenamefont
  {Hong},\ and\ \citenamefont {Gr{\"o}blacher}}]{riedinger_remote_2018}%
  \BibitemOpen
  \bibfield  {author} {\bibinfo {author} {\bibfnamefont {R.}~\bibnamefont
  {Riedinger}}, \bibinfo {author} {\bibfnamefont {A.}~\bibnamefont {Wallucks}},
  \bibinfo {author} {\bibfnamefont {I.}~\bibnamefont {Marinkovi{\'c}}},
  \bibinfo {author} {\bibfnamefont {C.}~\bibnamefont {L{\"o}schnauer}},
  \bibinfo {author} {\bibfnamefont {M.}~\bibnamefont {Aspelmeyer}}, \bibinfo
  {author} {\bibfnamefont {S.}~\bibnamefont {Hong}},\ and\ \bibinfo {author}
  {\bibfnamefont {S.}~\bibnamefont {Gr{\"o}blacher}},\ }``Remote quantum
  entanglement between two micromechanical oscillators'',\ \href
  {https://doi.org/10.1038/s41586-018-0036-z} {\bibfield  {journal} {\bibinfo
  {journal} {Nature}\ }\textbf {\bibinfo {volume} {556}},\ \bibinfo {pages}
  {473} (\bibinfo {year} {2018})}\BibitemShut {NoStop}%
\bibitem [{\citenamefont {Ockeloen-Korppi}\ \emph {et~al.}(2018)\citenamefont
  {Ockeloen-Korppi}, \citenamefont {Damsk{\"a}gg}, \citenamefont
  {Pirkkalainen}, \citenamefont {Asjad}, \citenamefont {Clerk}, \citenamefont
  {Massel}, \citenamefont {Woolley},\ and\ \citenamefont
  {Sillanp{\"a}{\"a}}}]{ockeloen-korppi_stabilized_2018}%
  \BibitemOpen
  \bibfield  {author} {\bibinfo {author} {\bibfnamefont {C.~F.}\ \bibnamefont
  {Ockeloen-Korppi}}, \bibinfo {author} {\bibfnamefont {E.}~\bibnamefont
  {Damsk{\"a}gg}}, \bibinfo {author} {\bibfnamefont {J.-M.}\ \bibnamefont
  {Pirkkalainen}}, \bibinfo {author} {\bibfnamefont {M.}~\bibnamefont {Asjad}},
  \bibinfo {author} {\bibfnamefont {A.~A.}\ \bibnamefont {Clerk}}, \bibinfo
  {author} {\bibfnamefont {F.}~\bibnamefont {Massel}}, \bibinfo {author}
  {\bibfnamefont {M.~J.}\ \bibnamefont {Woolley}},\ and\ \bibinfo {author}
  {\bibfnamefont {M.~A.}\ \bibnamefont {Sillanp{\"a}{\"a}}},\ }``Stabilized
  entanglement of massive mechanical oscillators'',\ \href
  {https://doi.org/10.1038/s41586-018-0038-x} {\bibfield  {journal} {\bibinfo
  {journal} {Nature}\ }\textbf {\bibinfo {volume} {556}},\ \bibinfo {pages}
  {478} (\bibinfo {year} {2018})}\BibitemShut {NoStop}%
\bibitem [{\citenamefont {Peterson}\ \emph {et~al.}(2019)\citenamefont
  {Peterson}, \citenamefont {Kotler}, \citenamefont {Lecocq}, \citenamefont
  {Cicak}, \citenamefont {Jin}, \citenamefont {Simmonds}, \citenamefont
  {Aumentado},\ and\ \citenamefont {Teufel}}]{peterson_ultrastrong_2019-1}%
  \BibitemOpen
  \bibfield  {author} {\bibinfo {author} {\bibfnamefont {G.~A.}\ \bibnamefont
  {Peterson}}, \bibinfo {author} {\bibfnamefont {S.}~\bibnamefont {Kotler}},
  \bibinfo {author} {\bibfnamefont {F.}~\bibnamefont {Lecocq}}, \bibinfo
  {author} {\bibfnamefont {K.}~\bibnamefont {Cicak}}, \bibinfo {author}
  {\bibfnamefont {X.~Y.}\ \bibnamefont {Jin}}, \bibinfo {author} {\bibfnamefont
  {R.~W.}\ \bibnamefont {Simmonds}}, \bibinfo {author} {\bibfnamefont
  {J.}~\bibnamefont {Aumentado}},\ and\ \bibinfo {author} {\bibfnamefont
  {J.~D.}\ \bibnamefont {Teufel}},\ }``Ultrastrong {Parametric} {Coupling}
  between a {Superconducting} {Cavity} and a {Mechanical} {Resonator}'',\ \href
  {https://doi.org/10.1103/PhysRevLett.123.247701} {\bibfield  {journal}
  {\bibinfo  {journal} {Physical Review Letters}\ }\textbf {\bibinfo {volume}
  {123}},\ \bibinfo {pages} {247701} (\bibinfo {year} {2019})}\BibitemShut
  {NoStop}%
\bibitem [{\citenamefont {Aspelmeyer}\ \emph {et~al.}(2014)\citenamefont
  {Aspelmeyer}, \citenamefont {Kippenberg},\ and\ \citenamefont
  {Marquardt}}]{aspelmeyer_cavity_2014}%
  \BibitemOpen
  \bibfield  {author} {\bibinfo {author} {\bibfnamefont {M.}~\bibnamefont
  {Aspelmeyer}}, \bibinfo {author} {\bibfnamefont {T.~J.}\ \bibnamefont
  {Kippenberg}},\ and\ \bibinfo {author} {\bibfnamefont {F.}~\bibnamefont
  {Marquardt}},\ }``Cavity optomechanics'',\ \href
  {https://doi.org/10.1103/RevModPhys.86.1391} {\bibfield  {journal} {\bibinfo
  {journal} {Reviews of Modern Physics}\ }\textbf {\bibinfo {volume} {86}},\
  \bibinfo {pages} {1391} (\bibinfo {year} {2014})}\BibitemShut {NoStop}%
\bibitem [{\citenamefont {Brennecke}\ \emph {et~al.}(2008)\citenamefont
  {Brennecke}, \citenamefont {Ritter}, \citenamefont {Donner},\ and\
  \citenamefont {Esslinger}}]{brennecke_cavity_2008}%
  \BibitemOpen
  \bibfield  {author} {\bibinfo {author} {\bibfnamefont {F.}~\bibnamefont
  {Brennecke}}, \bibinfo {author} {\bibfnamefont {S.}~\bibnamefont {Ritter}},
  \bibinfo {author} {\bibfnamefont {T.}~\bibnamefont {Donner}},\ and\ \bibinfo
  {author} {\bibfnamefont {T.}~\bibnamefont {Esslinger}},\ }``Cavity
  {Optomechanics} with a {Bose}-{Einstein} {Condensate}'',\ \href
  {https://doi.org/10.1126/science.1163218} {\bibfield  {journal} {\bibinfo
  {journal} {Science}\ }\textbf {\bibinfo {volume} {322}},\ \bibinfo {pages}
  {235} (\bibinfo {year} {2008})}\BibitemShut {NoStop}%
\bibitem [{\citenamefont {Murch}\ \emph {et~al.}(2008)\citenamefont {Murch},
  \citenamefont {Moore}, \citenamefont {Gupta},\ and\ \citenamefont
  {Stamper-Kurn}}]{murch_observation_2008}%
  \BibitemOpen
  \bibfield  {author} {\bibinfo {author} {\bibfnamefont {K.~W.}\ \bibnamefont
  {Murch}}, \bibinfo {author} {\bibfnamefont {K.~L.}\ \bibnamefont {Moore}},
  \bibinfo {author} {\bibfnamefont {S.}~\bibnamefont {Gupta}},\ and\ \bibinfo
  {author} {\bibfnamefont {D.~M.}\ \bibnamefont {Stamper-Kurn}},\
  }``Observation of quantum-measurement backaction with an ultracold atomic
  gas'',\ \href {https://doi.org/10.1038/nphys965} {\bibfield  {journal}
  {\bibinfo  {journal} {Nature Physics}\ }\textbf {\bibinfo {volume} {4}},\
  \bibinfo {pages} {561} (\bibinfo {year} {2008})}\BibitemShut {NoStop}%
\bibitem [{\citenamefont {Pirkkalainen}\ \emph
  {et~al.}(2015{\natexlab{b}})\citenamefont {Pirkkalainen}, \citenamefont
  {Cho}, \citenamefont {Massel}, \citenamefont {Tuorila}, \citenamefont
  {Heikkil{\"a}}, \citenamefont {Hakonen},\ and\ \citenamefont
  {Sillanp{\"a}{\"a}}}]{pirkkalainen_cavity_2015}%
  \BibitemOpen
  \bibfield  {author} {\bibinfo {author} {\bibfnamefont {J.-M.}\ \bibnamefont
  {Pirkkalainen}}, \bibinfo {author} {\bibfnamefont {S.~U.}\ \bibnamefont
  {Cho}}, \bibinfo {author} {\bibfnamefont {F.}~\bibnamefont {Massel}},
  \bibinfo {author} {\bibfnamefont {J.}~\bibnamefont {Tuorila}}, \bibinfo
  {author} {\bibfnamefont {T.~T.}\ \bibnamefont {Heikkil{\"a}}}, \bibinfo
  {author} {\bibfnamefont {P.~J.}\ \bibnamefont {Hakonen}},\ and\ \bibinfo
  {author} {\bibfnamefont {M.~A.}\ \bibnamefont {Sillanp{\"a}{\"a}}},\
  }``Cavity optomechanics mediated by a quantum two-level system'',\ \href
  {https://doi.org/10.1038/ncomms7981} {\bibfield  {journal} {\bibinfo
  {journal} {Nature Communications}\ }\textbf {\bibinfo {volume} {6}},\
  \bibinfo {pages} {1} (\bibinfo {year} {2015}{\natexlab{b}})}\BibitemShut
  {NoStop}%
\bibitem [{\citenamefont {Chu}\ \emph {et~al.}(2018)\citenamefont {Chu},
  \citenamefont {Kharel}, \citenamefont {Yoon}, \citenamefont {Frunzio},
  \citenamefont {Rakich},\ and\ \citenamefont
  {Schoelkopf}}]{chu_creation_2018}%
  \BibitemOpen
  \bibfield  {author} {\bibinfo {author} {\bibfnamefont {Y.}~\bibnamefont
  {Chu}}, \bibinfo {author} {\bibfnamefont {P.}~\bibnamefont {Kharel}},
  \bibinfo {author} {\bibfnamefont {T.}~\bibnamefont {Yoon}}, \bibinfo {author}
  {\bibfnamefont {L.}~\bibnamefont {Frunzio}}, \bibinfo {author} {\bibfnamefont
  {P.~T.}\ \bibnamefont {Rakich}},\ and\ \bibinfo {author} {\bibfnamefont
  {R.~J.}\ \bibnamefont {Schoelkopf}},\ }``Creation and control of multi-phonon
  {Fock} states in a bulk acoustic-wave resonator'',\ \href
  {https://doi.org/10.1038/s41586-018-0717-7} {\bibfield  {journal} {\bibinfo
  {journal} {Nature}\ }\textbf {\bibinfo {volume} {563}},\ \bibinfo {pages}
  {666} (\bibinfo {year} {2018})}\BibitemShut {NoStop}%
\bibitem [{\citenamefont {Viennot}\ \emph {et~al.}(2018)\citenamefont
  {Viennot}, \citenamefont {Ma},\ and\ \citenamefont
  {Lehnert}}]{viennot_phonon-number-sensitive_2018}%
  \BibitemOpen
  \bibfield  {author} {\bibinfo {author} {\bibfnamefont {J.~J.}\ \bibnamefont
  {Viennot}}, \bibinfo {author} {\bibfnamefont {X.}~\bibnamefont {Ma}},\ and\
  \bibinfo {author} {\bibfnamefont {K.~W.}\ \bibnamefont {Lehnert}},\
  }``Phonon-{Number}-{Sensitive} {Electromechanics}'',\ \bibfield  {journal}
  {\bibinfo  {journal} {Physical Review Letters}\ }\textbf {\bibinfo {volume}
  {121}},\ \bibinfo {pages} {183601} \href {https://doi.org/10.1103/PhysRevLett.121.183601}
(\bibinfo {year} {2018})\BibitemShut
  {NoStop}%
\bibitem [{\citenamefont {Delsing}\ \emph {et~al.}(2019)\citenamefont
  {Delsing}, \citenamefont {Cleland}, \citenamefont {Schuetz}, \citenamefont
  {Kn{\"o}rzer}, \citenamefont {Giedke}, \citenamefont {Cirac}, \citenamefont
  {Srinivasan}, \citenamefont {Wu}, \citenamefont {Balram}, \citenamefont
  {B{\"a}uerle}, \citenamefont {Meunier}, \citenamefont {Ford}, \citenamefont
  {Santos}, \citenamefont {Cerda-M{\'e}ndez}, \citenamefont {Wang},
  \citenamefont {Krenner}, \citenamefont {Nysten}, \citenamefont {Wei{\ss}},
  \citenamefont {Nash}, \citenamefont {Thevenard}, \citenamefont {Gourdon},
  \citenamefont {Rovillain}, \citenamefont {Marangolo}, \citenamefont
  {Duquesne}, \citenamefont {Fischerauer}, \citenamefont {Ruile}, \citenamefont
  {Reiner}, \citenamefont {Paschke}, \citenamefont {Denysenko}, \citenamefont
  {Volkmer}, \citenamefont {Wixforth}, \citenamefont {Bruus}, \citenamefont
  {Wiklund}, \citenamefont {Reboud}, \citenamefont {Cooper}, \citenamefont
  {Fu}, \citenamefont {Brugger}, \citenamefont {Rehfeldt},\ and\ \citenamefont
  {Westerhausen}}]{delsing_2019_2019}%
  \BibitemOpen
  \bibfield  {author} {\bibinfo {author} {\bibfnamefont {P.}~\bibnamefont
  {Delsing}}, \bibinfo {author} {\bibfnamefont {A.~N.}\ \bibnamefont
  {Cleland}}, \bibinfo {author} {\bibfnamefont {M.~J.~A.}\ \bibnamefont
  {Schuetz}}, \bibinfo {author} {\bibfnamefont {J.}~\bibnamefont
  {Kn{\"o}rzer}}, \bibinfo {author} {\bibfnamefont {G.}~\bibnamefont {Giedke}},
  \bibinfo {author} {\bibfnamefont {J.~I.}\ \bibnamefont {Cirac}}, \bibinfo
  {author} {\bibfnamefont {K.}~\bibnamefont {Srinivasan}}, \bibinfo {author}
  {\bibfnamefont {M.}~\bibnamefont {Wu}}, \bibinfo {author} {\bibfnamefont
  {K.~C.}\ \bibnamefont {Balram}}, \bibinfo {author} {\bibfnamefont
  {C.}~\bibnamefont {B{\"a}uerle}}, \bibinfo {author} {\bibfnamefont
  {T.}~\bibnamefont {Meunier}}, \bibinfo {author} {\bibfnamefont {C.~J.~B.}\
  \bibnamefont {Ford}}, \bibinfo {author} {\bibfnamefont {P.~V.}\ \bibnamefont
  {Santos}}, \bibinfo {author} {\bibfnamefont {E.}~\bibnamefont
  {Cerda-M{\'e}ndez}}, \bibinfo {author} {\bibfnamefont {H.}~\bibnamefont
  {Wang}}, \bibinfo {author} {\bibfnamefont {H.~J.}\ \bibnamefont {Krenner}},
  \bibinfo {author} {\bibfnamefont {E.~D.~S.}\ \bibnamefont {Nysten}}, \bibinfo
  {author} {\bibfnamefont {M.}~\bibnamefont {Wei{\ss}}}, \bibinfo {author}
  {\bibfnamefont {G.~R.}\ \bibnamefont {Nash}}, \bibinfo {author}
  {\bibfnamefont {L.}~\bibnamefont {Thevenard}}, \bibinfo {author}
  {\bibfnamefont {C.}~\bibnamefont {Gourdon}}, \bibinfo {author} {\bibfnamefont
  {P.}~\bibnamefont {Rovillain}}, \bibinfo {author} {\bibfnamefont
  {M.}~\bibnamefont {Marangolo}}, \bibinfo {author} {\bibfnamefont {J.~Y.}\
  \bibnamefont {Duquesne}}, \bibinfo {author} {\bibfnamefont {G.}~\bibnamefont
  {Fischerauer}}, \bibinfo {author} {\bibfnamefont {W.}~\bibnamefont {Ruile}},
  \bibinfo {author} {\bibfnamefont {A.}~\bibnamefont {Reiner}}, \bibinfo
  {author} {\bibfnamefont {B.}~\bibnamefont {Paschke}}, \bibinfo {author}
  {\bibfnamefont {D.}~\bibnamefont {Denysenko}}, \bibinfo {author}
  {\bibfnamefont {D.}~\bibnamefont {Volkmer}}, \bibinfo {author} {\bibfnamefont
  {A.}~\bibnamefont {Wixforth}}, \bibinfo {author} {\bibfnamefont
  {H.}~\bibnamefont {Bruus}}, \bibinfo {author} {\bibfnamefont
  {M.}~\bibnamefont {Wiklund}}, \bibinfo {author} {\bibfnamefont
  {J.}~\bibnamefont {Reboud}}, \bibinfo {author} {\bibfnamefont {J.~M.}\
  \bibnamefont {Cooper}}, \bibinfo {author} {\bibfnamefont {Y.}~\bibnamefont
  {Fu}}, \bibinfo {author} {\bibfnamefont {M.~S.}\ \bibnamefont {Brugger}},
  \bibinfo {author} {\bibfnamefont {F.}~\bibnamefont {Rehfeldt}},\ and\
  \bibinfo {author} {\bibfnamefont {C.}~\bibnamefont {Westerhausen}},\ }``The
  2019 surface acoustic waves roadmap'',\ \href
  {https://doi.org/10.1088/1361-6463/ab1b04} {\bibfield  {journal} {\bibinfo
  {journal} {Journal of Physics D: Applied Physics}\ }\textbf {\bibinfo
  {volume} {52}},\ \bibinfo {pages} {353001} (\bibinfo {year}
  {2019})}\BibitemShut {NoStop}%
\bibitem [{\citenamefont {Bera}\ \emph {et~al.}(2020)\citenamefont {Bera},
  \citenamefont {Majumder}, \citenamefont {Sahu},\ and\ \citenamefont
  {Singh}}]{bera_large_2020}%
  \BibitemOpen
  \bibfield  {author} {\bibinfo {author} {\bibfnamefont {T.}~\bibnamefont
  {Bera}}, \bibinfo {author} {\bibfnamefont {S.}~\bibnamefont {Majumder}},
  \bibinfo {author} {\bibfnamefont {S.~K.}\ \bibnamefont {Sahu}},\ and\
  \bibinfo {author} {\bibfnamefont {V.}~\bibnamefont {Singh}},\ }``Large
  flux-mediated coupling in hybrid electromechanical system with a transmon
  qubit'',\ \href {https://doi.org/10.1038/s42005-020-00514-y} {\bibfield  {journal}
  {\bibinfo  {journal} {Communications Physics}\ } \textbf {\bibinfo {volume}
  {4}},\ \bibinfo {pages} {12} (\bibinfo {year}
  {2020})}\BibitemShut {NoStop}%
\bibitem [{\citenamefont {Arndt}\ and\ \citenamefont
  {Hornberger}(2014)}]{arndt_testing_2014}%
  \BibitemOpen
  \bibfield  {author} {\bibinfo {author} {\bibfnamefont {M.}~\bibnamefont
  {Arndt}}\ and\ \bibinfo {author} {\bibfnamefont {K.}~\bibnamefont
  {Hornberger}},\ }``Testing the limits of quantum mechanical
  superpositions'',\ \href {https://doi.org/10.1038/nphys2863} {\bibfield
  {journal} {\bibinfo  {journal} {Nature Physics}\ }\textbf {\bibinfo {volume}
  {10}},\ \bibinfo {pages} {271} (\bibinfo {year} {2014})}\BibitemShut
  {NoStop}%
\bibitem [{\citenamefont {Regal}\ and\ \citenamefont
  {Lehnert}(2011)}]{regal_cavity_2011}%
  \BibitemOpen
  \bibfield  {author} {\bibinfo {author} {\bibfnamefont {C.~A.}\ \bibnamefont
  {Regal}}\ and\ \bibinfo {author} {\bibfnamefont {K.~W.}\ \bibnamefont
  {Lehnert}},\ }``From cavity electromechanics to cavity optomechanics'',\
  \href {https://doi.org/10.1088/1742-6596/264/1/012025} {\bibfield  {journal}
  {\bibinfo  {journal} {Journal of Physics: Conference Series}\ }\textbf
  {\bibinfo {volume} {264}},\ \bibinfo {pages} {012025} (\bibinfo {year}
  {2011})}\BibitemShut {NoStop}%
\bibitem [{\citenamefont {Yuan}\ \emph
  {et~al.}(2015{\natexlab{a}})\citenamefont {Yuan}, \citenamefont {Singh},
  \citenamefont {Blanter},\ and\ \citenamefont {Steele}}]{yuan_large_2015}%
  \BibitemOpen
  \bibfield  {author} {\bibinfo {author} {\bibfnamefont {M.}~\bibnamefont
  {Yuan}}, \bibinfo {author} {\bibfnamefont {V.}~\bibnamefont {Singh}},
  \bibinfo {author} {\bibfnamefont {Y.~M.}\ \bibnamefont {Blanter}},\ and\
  \bibinfo {author} {\bibfnamefont {G.~A.}\ \bibnamefont {Steele}},\ }``Large
  cooperativity and microkelvin cooling with a three-dimensional optomechanical
  cavity'',\ \bibfield  {journal} {\bibinfo  {journal} {Nature Communications}\
  }\textbf {\bibinfo {volume} {6}},\ \bibinfo {pages} {8491} \href {https://doi.org/10.1038/ncomms9491}
(\bibinfo {year} {2015}{\natexlab{a}})\BibitemShut
  {NoStop}%
\bibitem [{\citenamefont {Guo}\ \emph {et~al.}(2019)\citenamefont {Guo},
  \citenamefont {Norte},\ and\ \citenamefont
  {Gr{\"o}blacher}}]{guo_feedback_2019}%
  \BibitemOpen
  \bibfield  {author} {\bibinfo {author} {\bibfnamefont {J.}~\bibnamefont
  {Guo}}, \bibinfo {author} {\bibfnamefont {R.}~\bibnamefont {Norte}},\ and\
  \bibinfo {author} {\bibfnamefont {S.}~\bibnamefont {Gr{\"o}blacher}},\
  }``Feedback {Cooling} of a {Room} {Temperature} {Mechanical} {Oscillator}
  close to its {Motional} {Ground} {State}'',\ \href
  {https://doi.org/10.1103/PhysRevLett.123.223602} {\bibfield  {journal}
  {\bibinfo  {journal} {Physical Review Letters}\ }\textbf {\bibinfo {volume}
  {123}},\ \bibinfo {pages} {223602} (\bibinfo {year} {2019})}\BibitemShut
  {NoStop}%
\bibitem [{noa()}]{noauthor_see_nodate}%
  \BibitemOpen
  See supplementary material which includes {Refs}. [14], [31-38],
  [41,42] for additional information on the sample preparation, the measurement
  setup and characterisation measurements, the measurement protocol and data
  treatment, data on the temperature dependence of the device parameters,
  dependence of the cooperativity on the cavity bias point and estimations of
  the maximum photon number\BibitemShut {NoStop}%
\bibitem [{\citenamefont {Via}\ \emph {et~al.}(2015)\citenamefont {Via},
  \citenamefont {Kirchmair},\ and\ \citenamefont
  {Romero-Isart}}]{via_strong_2015}%
  \BibitemOpen
  \bibfield  {author} {\bibinfo {author} {\bibfnamefont {G.}~\bibnamefont
  {Via}}, \bibinfo {author} {\bibfnamefont {G.}~\bibnamefont {Kirchmair}},\
  and\ \bibinfo {author} {\bibfnamefont {O.}~\bibnamefont {Romero-Isart}},\
  }``Strong {Single}-{Photon} {Coupling} in {Superconducting} {Quantum}
  {Magnetomechanics}'',\ \bibfield  {journal} {\bibinfo  {journal} {Physical
  Review Letters}\ }\textbf {\bibinfo {volume} {114}},\ \bibinfo {pages} {143602} \href
  {https://doi.org/10.1103/PhysRevLett.114.143602}
(\bibinfo {year} {2015})\BibitemShut
  {NoStop}%
\bibitem [{\citenamefont {Rodrigues}\ \emph {et~al.}(2019)\citenamefont
  {Rodrigues}, \citenamefont {Bothner},\ and\ \citenamefont
  {Steele}}]{rodrigues_coupling_2019}%
  \BibitemOpen
  \bibfield  {author} {\bibinfo {author} {\bibfnamefont {I.~C.}\ \bibnamefont
  {Rodrigues}}, \bibinfo {author} {\bibfnamefont {D.}~\bibnamefont {Bothner}},\
  and\ \bibinfo {author} {\bibfnamefont {G.~A.}\ \bibnamefont {Steele}},\
  }``Coupling microwave photons to a mechanical resonator using quantum
  interference'',\ \href {https://doi.org/10.1038/s41467-019-12964-2}
  {\bibfield  {journal} {\bibinfo  {journal} {Nature Communications}\ }\textbf
  {\bibinfo {volume} {10}},\ \bibinfo {pages} {1} (\bibinfo {year}
  {2019})}\BibitemShut {NoStop}%
\bibitem [{\citenamefont {Wang}\ \emph {et~al.}(2014)\citenamefont {Wang},
  \citenamefont {Li}, \citenamefont {Li}, \citenamefont {Jiang}, \citenamefont
  {Gao},\ and\ \citenamefont {Li}}]{wang_preparing_2014}%
  \BibitemOpen
  \bibfield  {author} {\bibinfo {author} {\bibfnamefont {X.}~\bibnamefont
  {Wang}}, \bibinfo {author} {\bibfnamefont {H.~R.}\ \bibnamefont {Li}},
  \bibinfo {author} {\bibfnamefont {P.~B.}\ \bibnamefont {Li}}, \bibinfo
  {author} {\bibfnamefont {C.~W.}\ \bibnamefont {Jiang}}, \bibinfo {author}
  {\bibfnamefont {H.}~\bibnamefont {Gao}},\ and\ \bibinfo {author}
  {\bibfnamefont {F.~L.}\ \bibnamefont {Li}},\ }``Preparing ground states and
  squeezed states of nanomechanical cantilevers by fast dissipation'',\ \href
  {https://doi.org/10.1103/PhysRevA.90.013838} {\bibfield  {journal} {\bibinfo
  {journal} {Physical Review A}\ }\textbf {\bibinfo {volume} {90}},\ \bibinfo
  {pages} {013838} (\bibinfo {year} {2014})}\BibitemShut {NoStop}%
\bibitem [{\citenamefont {Schmidt}\ \emph {et~al.}(2019)\citenamefont
  {Schmidt}, \citenamefont {Amawi}, \citenamefont {Pogorzalek}, \citenamefont
  {Deppe}, \citenamefont {Marx}, \citenamefont {Gross},\ and\ \citenamefont
  {Huebl}}]{schmidt_sideband-resolved_2019}%
  \BibitemOpen
  \bibfield  {author} {\bibinfo {author} {\bibfnamefont {P.}~\bibnamefont
  {Schmidt}}, \bibinfo {author} {\bibfnamefont {M.~T.}\ \bibnamefont {Amawi}},
  \bibinfo {author} {\bibfnamefont {S.}~\bibnamefont {Pogorzalek}}, \bibinfo
  {author} {\bibfnamefont {F.}~\bibnamefont {Deppe}}, \bibinfo {author}
  {\bibfnamefont {A.}~\bibnamefont {Marx}}, \bibinfo {author} {\bibfnamefont
  {R.}~\bibnamefont {Gross}},\ and\ \bibinfo {author} {\bibfnamefont
  {H.}~\bibnamefont {Huebl}},\ }``Sideband-resolved resonator electromechanics 
  based on a nonlinear Josephson inductance probed on the single-photon level'',\
  \href {https://doi.org/10.1038/s42005-020-00501-3} {\bibfield  {journal} {\bibinfo
  {journal} {Nature Communications}\ }\textbf {\bibinfo {volume} {3}},\ \bibinfo
  {pages} {233} (\bibinfo {year} {2020})}\BibitemShut
  {NoStop}%
\bibitem [{\citenamefont {Zoepfl}\ \emph {et~al.}(2017)\citenamefont {Zoepfl},
  \citenamefont {Muppalla}, \citenamefont {Schneider}, \citenamefont
  {Kasemann}, \citenamefont {Partel},\ and\ \citenamefont
  {Kirchmair}}]{zoepfl_characterization_2017}%
  \BibitemOpen
  \bibfield  {author} {\bibinfo {author} {\bibfnamefont {D.}~\bibnamefont
  {Zoepfl}}, \bibinfo {author} {\bibfnamefont {P.~R.}\ \bibnamefont
  {Muppalla}}, \bibinfo {author} {\bibfnamefont {C.~M.~F.}\ \bibnamefont
  {Schneider}}, \bibinfo {author} {\bibfnamefont {S.}~\bibnamefont {Kasemann}},
  \bibinfo {author} {\bibfnamefont {S.}~\bibnamefont {Partel}},\ and\ \bibinfo
  {author} {\bibfnamefont {G.}~\bibnamefont {Kirchmair}},\ }``Characterization
  of low loss microstrip resonators as a building block for circuit {QED} in a
  3D waveguide'',\ \href {https://doi.org/10.1063/1.4992070} {\bibfield
  {journal} {\bibinfo  {journal} {AIP Advances}\ }\textbf {\bibinfo {volume}
  {7}},\ \bibinfo {pages} {085118} (\bibinfo {year} {2017})}\BibitemShut
  {NoStop}%
\bibitem [{\citenamefont {Probst}\ \emph {et~al.}(2015)\citenamefont {Probst},
  \citenamefont {Song}, \citenamefont {Bushev}, \citenamefont {Ustinov},\ and\
  \citenamefont {Weides}}]{probst_efficient_2015}%
  \BibitemOpen
  \bibfield  {author} {\bibinfo {author} {\bibfnamefont {S.}~\bibnamefont
  {Probst}}, \bibinfo {author} {\bibfnamefont {F.~B.}\ \bibnamefont {Song}},
  \bibinfo {author} {\bibfnamefont {P.~A.}\ \bibnamefont {Bushev}}, \bibinfo
  {author} {\bibfnamefont {A.~V.}\ \bibnamefont {Ustinov}},\ and\ \bibinfo
  {author} {\bibfnamefont {M.}~\bibnamefont {Weides}},\ }``Efficient and robust
  analysis of complex scattering data under noise in microwave resonators'',\
  \href {https://doi.org/10.1063/1.4907935} {\bibfield  {journal} {\bibinfo
  {journal} {Review of Scientific Instruments}\ }\textbf {\bibinfo {volume}
  {86}},\ \bibinfo {pages} {024706} (\bibinfo {year} {2015})}\BibitemShut
  {NoStop}%
\bibitem [{\citenamefont {Khalil}\ \emph {et~al.}(2012)\citenamefont {Khalil},
  \citenamefont {Stoutimore}, \citenamefont {Wellstood},\ and\ \citenamefont
  {Osborn}}]{khalil_analysis_2012}%
  \BibitemOpen
  \bibfield  {author} {\bibinfo {author} {\bibfnamefont {M.~S.}\ \bibnamefont
  {Khalil}}, \bibinfo {author} {\bibfnamefont {M.~J.~A.}\ \bibnamefont
  {Stoutimore}}, \bibinfo {author} {\bibfnamefont {F.~C.}\ \bibnamefont
  {Wellstood}},\ and\ \bibinfo {author} {\bibfnamefont {K.~D.}\ \bibnamefont
  {Osborn}},\ }``An analysis method for asymmetric resonator transmission
  applied to superconducting devices'',\ \href
  {https://doi.org/10.1063/1.3692073} {\bibfield  {journal} {\bibinfo
  {journal} {Journal of Applied Physics}\ }\textbf {\bibinfo {volume} {111}},\
  \bibinfo {pages} {054510} (\bibinfo {year} {2012})}\BibitemShut {NoStop}%
\bibitem [{\citenamefont {Gorodetksy}\ \emph {et~al.}(2010)\citenamefont
  {Gorodetksy}, \citenamefont {Schliesser}, \citenamefont {Anetsberger},
  \citenamefont {Deleglise},\ and\ \citenamefont
  {Kippenberg}}]{gorodetksy_determination_2010}%
  \BibitemOpen
  \bibfield  {author} {\bibinfo {author} {\bibfnamefont {M.~L.}\ \bibnamefont
  {Gorodetksy}}, \bibinfo {author} {\bibfnamefont {A.}~\bibnamefont
  {Schliesser}}, \bibinfo {author} {\bibfnamefont {G.}~\bibnamefont
  {Anetsberger}}, \bibinfo {author} {\bibfnamefont {S.}~\bibnamefont
  {Deleglise}},\ and\ \bibinfo {author} {\bibfnamefont {T.~J.}\ \bibnamefont
  {Kippenberg}},\ }``Determination of the vacuum optomechanical coupling rate
  using frequency noise calibration'',\ \href
  {https://doi.org/10.1364/OE.18.023236} {\bibfield  {journal} {\bibinfo
  {journal} {Optics Express}\ }\textbf {\bibinfo {volume} {18}},\ \bibinfo
  {pages} {23236} (\bibinfo {year} {2010})}\BibitemShut {NoStop}%
\bibitem [{\citenamefont {Zhou}\ \emph {et~al.}(2013)\citenamefont {Zhou},
  \citenamefont {Hocke}, \citenamefont {Schliesser}, \citenamefont {Marx},
  \citenamefont {Huebl}, \citenamefont {Gross},\ and\ \citenamefont
  {Kippenberg}}]{zhou_slowing_2013}%
  \BibitemOpen
  \bibfield  {author} {\bibinfo {author} {\bibfnamefont {X.}~\bibnamefont
  {Zhou}}, \bibinfo {author} {\bibfnamefont {F.}~\bibnamefont {Hocke}},
  \bibinfo {author} {\bibfnamefont {A.}~\bibnamefont {Schliesser}}, \bibinfo
  {author} {\bibfnamefont {A.}~\bibnamefont {Marx}}, \bibinfo {author}
  {\bibfnamefont {H.}~\bibnamefont {Huebl}}, \bibinfo {author} {\bibfnamefont
  {R.}~\bibnamefont {Gross}},\ and\ \bibinfo {author} {\bibfnamefont {T.~J.}\
  \bibnamefont {Kippenberg}},\ }``Slowing, advancing and switching of microwave
  signals using circuit nanoelectromechanics'',\ \href
  {https://doi.org/10.1038/nphys2527} {\bibfield  {journal} {\bibinfo
  {journal} {Nature Physics}\ }\textbf {\bibinfo {volume} {9}},\ \bibinfo
  {pages} {179} (\bibinfo {year} {2013})}\BibitemShut {NoStop}%
\bibitem [{\citenamefont {Marquardt}\ \emph {et~al.}(2007)\citenamefont
  {Marquardt}, \citenamefont {Chen}, \citenamefont {Clerk},\ and\ \citenamefont
  {Girvin}}]{marquardt_quantum_2007}%
  \BibitemOpen
  \bibfield  {author} {\bibinfo {author} {\bibfnamefont {F.}~\bibnamefont
  {Marquardt}}, \bibinfo {author} {\bibfnamefont {J.~P.}\ \bibnamefont {Chen}},
  \bibinfo {author} {\bibfnamefont {A.~A.}\ \bibnamefont {Clerk}},\ and\
  \bibinfo {author} {\bibfnamefont {S.~M.}\ \bibnamefont {Girvin}},\ }``Quantum
  {Theory} of {Cavity}-{Assisted} {Sideband} {Cooling} of {Mechanical}
  {Motion}'',\ \bibfield  {journal} {\bibinfo  {journal} {Physical Review
  Letters}\ }\textbf {\bibinfo {volume} {99}},\ \bibinfo{pages} {093902} \href
  {https://doi.org/10.1103/PhysRevLett.99.093902}
(\bibinfo {year} {2007})\BibitemShut
  {NoStop}%
\bibitem [{\citenamefont {Safavi-Naeini}\ \emph {et~al.}(2013)\citenamefont
  {Safavi-Naeini}, \citenamefont {Chan}, \citenamefont {Hill}, \citenamefont
  {Gr{\"o}blacher}, \citenamefont {Miao}, \citenamefont {Chen}, \citenamefont
  {Aspelmeyer},\ and\ \citenamefont {Painter}}]{safavi-naeini_laser_2013}%
  \BibitemOpen
  \bibfield  {author} {\bibinfo {author} {\bibfnamefont {A.~H.}\ \bibnamefont
  {Safavi-Naeini}}, \bibinfo {author} {\bibfnamefont {J.}~\bibnamefont {Chan}},
  \bibinfo {author} {\bibfnamefont {J.~T.}\ \bibnamefont {Hill}}, \bibinfo
  {author} {\bibfnamefont {S.}~\bibnamefont {Gr{\"o}blacher}}, \bibinfo
  {author} {\bibfnamefont {H.}~\bibnamefont {Miao}}, \bibinfo {author}
  {\bibfnamefont {Y.}~\bibnamefont {Chen}}, \bibinfo {author} {\bibfnamefont
  {M.}~\bibnamefont {Aspelmeyer}},\ and\ \bibinfo {author} {\bibfnamefont
  {O.}~\bibnamefont {Painter}},\ }``Laser noise in cavity-optomechanical
  cooling and thermometry'',\ \href
  {https://doi.org/10.1088/1367-2630/15/3/035007} {\bibfield  {journal}
  {\bibinfo  {journal} {New Journal of Physics}\ }\textbf {\bibinfo {volume}
  {15}},\ \bibinfo {pages} {035007} (\bibinfo {year} {2013})}\BibitemShut
  {NoStop}%
\bibitem [{\citenamefont {Nation}\ \emph {et~al.}(2008)\citenamefont {Nation},
  \citenamefont {Blencowe},\ and\ \citenamefont {Buks}}]{nation_quantum_2008}%
  \BibitemOpen
  \bibfield  {author} {\bibinfo {author} {\bibfnamefont {P.~D.}\ \bibnamefont
  {Nation}}, \bibinfo {author} {\bibfnamefont {M.~P.}\ \bibnamefont
  {Blencowe}},\ and\ \bibinfo {author} {\bibfnamefont {E.}~\bibnamefont
  {Buks}},\ }``Quantum analysis of a nonlinear microwave cavity-embedded dc
  {SQUID} displacement detector'',\ \href
  {https://doi.org/10.1103/PhysRevB.78.104516} {\bibfield  {journal} {\bibinfo
  {journal} {Physical Review B}\ }\textbf {\bibinfo {volume} {78}},\ \bibinfo
  {pages} {104516} (\bibinfo {year} {2008})}\BibitemShut {NoStop}%
\bibitem [{\citenamefont {Mamin}\ and\ \citenamefont
  {Rugar}(2001)}]{mamin_sub-attonewton_2001}%
  \BibitemOpen
  \bibfield  {author} {\bibinfo {author} {\bibfnamefont {H.~J.}\ \bibnamefont
  {Mamin}}\ and\ \bibinfo {author} {\bibfnamefont {D.}~\bibnamefont {Rugar}},\
  }``Sub-attonewton force detection at millikelvin temperatures'',\ \href
  {https://doi.org/10.1063/1.1418256} {\bibfield  {journal} {\bibinfo
  {journal} {Applied Physics Letters}\ }\textbf {\bibinfo {volume} {79}},\
  \bibinfo {pages} {3358} (\bibinfo {year} {2001})}\BibitemShut {NoStop}%
\bibitem [{\citenamefont {Andrews}\ \emph {et~al.}(2014)\citenamefont
  {Andrews}, \citenamefont {Peterson}, \citenamefont {Purdy}, \citenamefont
  {Cicak}, \citenamefont {Simmonds}, \citenamefont {Regal},\ and\ \citenamefont
  {Lehnert}}]{andrews_bidirectional_2014}%
  \BibitemOpen
  \bibfield  {author} {\bibinfo {author} {\bibfnamefont {R.~W.}\ \bibnamefont
  {Andrews}}, \bibinfo {author} {\bibfnamefont {R.~W.}\ \bibnamefont
  {Peterson}}, \bibinfo {author} {\bibfnamefont {T.~P.}\ \bibnamefont {Purdy}},
  \bibinfo {author} {\bibfnamefont {K.}~\bibnamefont {Cicak}}, \bibinfo
  {author} {\bibfnamefont {R.~W.}\ \bibnamefont {Simmonds}}, \bibinfo {author}
  {\bibfnamefont {C.~A.}\ \bibnamefont {Regal}},\ and\ \bibinfo {author}
  {\bibfnamefont {K.~W.}\ \bibnamefont {Lehnert}},\ }``Bidirectional and
  efficient conversion between microwave and optical light'',\ \href
  {https://doi.org/10.1038/nphys2911} {\bibfield  {journal} {\bibinfo
  {journal} {Nature Physics}\ }\textbf {\bibinfo {volume} {10}},\ \bibinfo
  {pages} {321} (\bibinfo {year} {2014})}\BibitemShut {NoStop}%
\bibitem [{\citenamefont {Yuan}\ \emph
  {et~al.}(2015{\natexlab{b}})\citenamefont {Yuan}, \citenamefont {Cohen},\
  and\ \citenamefont {Steele}}]{yuan_silicon_2015}%
  \BibitemOpen
  \bibfield  {author} {\bibinfo {author} {\bibfnamefont {M.}~\bibnamefont
  {Yuan}}, \bibinfo {author} {\bibfnamefont {M.~A.}\ \bibnamefont {Cohen}},\
  and\ \bibinfo {author} {\bibfnamefont {G.~A.}\ \bibnamefont {Steele}},\
  }``Silicon nitride membrane resonators at millikelvin temperatures with
  quality factors exceeding 108'',\ \href {https://doi.org/10.1063/1.4938747}
  {\bibfield  {journal} {\bibinfo  {journal} {Applied Physics Letters}\
  }\textbf {\bibinfo {volume} {107}},\ \bibinfo {pages} {263501} (\bibinfo
  {year} {2015}{\natexlab{b}})}\BibitemShut {NoStop}%
\bibitem [{\citenamefont {Agrawal}\ and\ \citenamefont
  {Carmichael}(1979)}]{agrawal_optical_1979}%
  \BibitemOpen
  \bibfield  {author} {\bibinfo {author} {\bibfnamefont {G.~P.}\ \bibnamefont
  {Agrawal}}\ and\ \bibinfo {author} {\bibfnamefont {H.~J.}\ \bibnamefont
  {Carmichael}},\ }``Optical bistability through nonlinear dispersion and
  absorption'',\ \href {https://doi.org/10.1103/PhysRevA.19.2074} {\bibfield
  {journal} {\bibinfo  {journal} {Physical Review A}\ }\textbf {\bibinfo
  {volume} {19}},\ \bibinfo {pages} {2074} (\bibinfo {year}
  {1979})}\BibitemShut {NoStop}%
\end{thebibliography}

\begin{thebibliography}{11}%
\makeatletter
\providecommand \@ifxundefined [1]{%
 \@ifx{#1\undefined}
}%
\providecommand \@ifnum [1]{%
 \ifnum #1\expandafter \@firstoftwo
 \else \expandafter \@secondoftwo
 \fi
}%
\providecommand \@ifx [1]{%
 \ifx #1\expandafter \@firstoftwo
 \else \expandafter \@secondoftwo
 \fi
}%
\providecommand \natexlab [1]{#1}%
\providecommand \enquote  [1]{``#1''}%
\providecommand \bibnamefont  [1]{#1}%
\providecommand \bibfnamefont [1]{#1}%
\providecommand \citenamefont [1]{#1}%
\providecommand \href@noop [0]{\@secondoftwo}%
\providecommand \href [0]{\begingroup \@sanitize@url \@href}%
\providecommand \@href[1]{\@@startlink{#1}\@@href}%
\providecommand \@@href[1]{\endgroup#1\@@endlink}%
\providecommand \@sanitize@url [0]{\catcode `\\12\catcode `\$12\catcode
  `\&12\catcode `\#12\catcode `\^12\catcode `\_12\catcode `\%12\relax}%
\providecommand \@@startlink[1]{}%
\providecommand \@@endlink[0]{}%
\providecommand \url  [0]{\begingroup\@sanitize@url \@url }%
\providecommand \@url [1]{\endgroup\@href {#1}{\urlprefix }}%
\providecommand \urlprefix  [0]{URL }%
\providecommand \Eprint [0]{\href }%
\providecommand \doibase [0]{https://doi.org/}%
\providecommand \selectlanguage [0]{\@gobble}%
\providecommand \bibinfo  [0]{\@secondoftwo}%
\providecommand \bibfield  [0]{\@secondoftwo}%
\providecommand \translation [1]{[#1]}%
\providecommand \BibitemOpen [0]{}%
\providecommand \bibitemStop [0]{}%
\providecommand \bibitemNoStop [0]{.\EOS\space}%
\providecommand \EOS [0]{\spacefactor3000\relax}%
\providecommand \BibitemShut  [1]{\csname bibitem#1\endcsname}%
\let\auto@bib@innerbib\@empty
\bibitem [{\citenamefont {Aspelmeyer}\ \emph {et~al.}(2014)\citenamefont
  {Aspelmeyer}, \citenamefont {Kippenberg},\ and\ \citenamefont
  {Marquardt}}]{Saspelmeyer_cavity_2014}%
  \BibitemOpen
  \bibfield  {author} {\bibinfo {author} {\bibfnamefont {M.}~\bibnamefont
  {Aspelmeyer}}, \bibinfo {author} {\bibfnamefont {T.~J.}\ \bibnamefont
  {Kippenberg}},\ and\ \bibinfo {author} {\bibfnamefont {F.}~\bibnamefont
  {Marquardt}},\ }``Cavity optomechanics'',\ \href
  {https://doi.org/10.1103/RevModPhys.86.1391} {\bibfield  {journal} {\bibinfo
  {journal} {Reviews of Modern Physics}\ }\textbf {\bibinfo {volume} {86}},\
  \bibinfo {pages} {1391} (\bibinfo {year} {2014})}\BibitemShut {NoStop}%
\bibitem [{\citenamefont {Zoepfl}\ \emph {et~al.}(2017)\citenamefont {Zoepfl},
  \citenamefont {Muppalla}, \citenamefont {Schneider}, \citenamefont
  {Kasemann}, \citenamefont {Partel},\ and\ \citenamefont
  {Kirchmair}}]{Szoepfl_characterization_2017}%
  \BibitemOpen
  \bibfield  {author} {\bibinfo {author} {\bibfnamefont {D.}~\bibnamefont
  {Zoepfl}}, \bibinfo {author} {\bibfnamefont {P.~R.}\ \bibnamefont
  {Muppalla}}, \bibinfo {author} {\bibfnamefont {C.~M.~F.}\ \bibnamefont
  {Schneider}}, \bibinfo {author} {\bibfnamefont {S.}~\bibnamefont {Kasemann}},
  \bibinfo {author} {\bibfnamefont {S.}~\bibnamefont {Partel}},\ and\ \bibinfo
  {author} {\bibfnamefont {G.}~\bibnamefont {Kirchmair}},\ }``Characterization
  of low loss microstrip resonators as a building block for circuit {QED} in a
  3d waveguide'',\ \href {https://doi.org/10.1063/1.4992070} {\bibfield
  {journal} {\bibinfo  {journal} {AIP Advances}\ }\textbf {\bibinfo {volume}
  {7}},\ \bibinfo {pages} {085118} (\bibinfo {year} {2017})}\BibitemShut
  {NoStop}%
\bibitem [{\citenamefont {Probst}\ \emph {et~al.}(2015)\citenamefont {Probst},
  \citenamefont {Song}, \citenamefont {Bushev}, \citenamefont {Ustinov},\ and\
  \citenamefont {Weides}}]{Sprobst_efficient_2015}%
  \BibitemOpen
  \bibfield  {author} {\bibinfo {author} {\bibfnamefont {S.}~\bibnamefont
  {Probst}}, \bibinfo {author} {\bibfnamefont {F.~B.}\ \bibnamefont {Song}},
  \bibinfo {author} {\bibfnamefont {P.~A.}\ \bibnamefont {Bushev}}, \bibinfo
  {author} {\bibfnamefont {A.~V.}\ \bibnamefont {Ustinov}},\ and\ \bibinfo
  {author} {\bibfnamefont {M.}~\bibnamefont {Weides}},\ }``Efficient and robust
  analysis of complex scattering data under noise in microwave resonators'',\
  \href {https://doi.org/10.1063/1.4907935} {\bibfield  {journal} {\bibinfo
  {journal} {Review of Scientific Instruments}\ }\textbf {\bibinfo {volume}
  {86}},\ \bibinfo {pages} {024706} (\bibinfo {year} {2015})}\BibitemShut
  {NoStop}%
\bibitem [{\citenamefont {Khalil}\ \emph {et~al.}(2012)\citenamefont {Khalil},
  \citenamefont {Stoutimore}, \citenamefont {Wellstood},\ and\ \citenamefont
  {Osborn}}]{Skhalil_analysis_2012}%
  \BibitemOpen
  \bibfield  {author} {\bibinfo {author} {\bibfnamefont {M.~S.}\ \bibnamefont
  {Khalil}}, \bibinfo {author} {\bibfnamefont {M.~J.~A.}\ \bibnamefont
  {Stoutimore}}, \bibinfo {author} {\bibfnamefont {F.~C.}\ \bibnamefont
  {Wellstood}},\ and\ \bibinfo {author} {\bibfnamefont {K.~D.}\ \bibnamefont
  {Osborn}},\ }``An analysis method for asymmetric resonator transmission
  applied to superconducting devices'',\ \href
  {https://doi.org/10.1063/1.3692073} {\bibfield  {journal} {\bibinfo
  {journal} {Journal of Applied Physics}\ }\textbf {\bibinfo {volume} {111}},\
  \bibinfo {pages} {054510} (\bibinfo {year} {2012})}\BibitemShut {NoStop}%
\bibitem [{\citenamefont {Gorodetksy}\ \emph {et~al.}(2010)\citenamefont
  {Gorodetksy}, \citenamefont {Schliesser}, \citenamefont {Anetsberger},
  \citenamefont {Deleglise},\ and\ \citenamefont
  {Kippenberg}}]{Sgorodetksy_determination_2010}%
  \BibitemOpen
  \bibfield  {author} {\bibinfo {author} {\bibfnamefont {M.~L.}\ \bibnamefont
  {Gorodetksy}}, \bibinfo {author} {\bibfnamefont {A.}~\bibnamefont
  {Schliesser}}, \bibinfo {author} {\bibfnamefont {G.}~\bibnamefont
  {Anetsberger}}, \bibinfo {author} {\bibfnamefont {S.}~\bibnamefont
  {Deleglise}},\ and\ \bibinfo {author} {\bibfnamefont {T.~J.}\ \bibnamefont
  {Kippenberg}},\ }``Determination of the vacuum optomechanical coupling rate
  using frequency noise calibration'',\ \href
  {https://doi.org/10.1364/OE.18.023236} {\bibfield  {journal} {\bibinfo
  {journal} {Optics Express}\ }\textbf {\bibinfo {volume} {18}},\ \bibinfo
  {pages} {23236} (\bibinfo {year} {2010})}\BibitemShut {NoStop}%
\bibitem [{\citenamefont {Zhou}\ \emph {et~al.}(2013)\citenamefont {Zhou},
  \citenamefont {Hocke}, \citenamefont {Schliesser}, \citenamefont {Marx},
  \citenamefont {Huebl}, \citenamefont {Gross},\ and\ \citenamefont
  {Kippenberg}}]{Szhou_slowing_2013}%
  \BibitemOpen
  \bibfield  {author} {\bibinfo {author} {\bibfnamefont {X.}~\bibnamefont
  {Zhou}}, \bibinfo {author} {\bibfnamefont {F.}~\bibnamefont {Hocke}},
  \bibinfo {author} {\bibfnamefont {A.}~\bibnamefont {Schliesser}}, \bibinfo
  {author} {\bibfnamefont {A.}~\bibnamefont {Marx}}, \bibinfo {author}
  {\bibfnamefont {H.}~\bibnamefont {Huebl}}, \bibinfo {author} {\bibfnamefont
  {R.}~\bibnamefont {Gross}},\ and\ \bibinfo {author} {\bibfnamefont {T.~J.}\
  \bibnamefont {Kippenberg}},\ }``Slowing, advancing and switching of microwave
  signals using circuit nanoelectromechanics'',\ \href
  {https://doi.org/10.1038/nphys2527} {\bibfield  {journal} {\bibinfo
  {journal} {Nature Physics}\ }\textbf {\bibinfo {volume} {9}},\ \bibinfo
  {pages} {179} (\bibinfo {year} {2013})}\BibitemShut {NoStop}%
\bibitem [{\citenamefont {Marquardt}\ \emph {et~al.}(2007)\citenamefont
  {Marquardt}, \citenamefont {Chen}, \citenamefont {Clerk},\ and\ \citenamefont
  {Girvin}}]{Smarquardt_quantum_2007}%
  \BibitemOpen
  \bibfield  {author} {\bibinfo {author} {\bibfnamefont {F.}~\bibnamefont
  {Marquardt}}, \bibinfo {author} {\bibfnamefont {J.~P.}\ \bibnamefont {Chen}},
  \bibinfo {author} {\bibfnamefont {A.~A.}\ \bibnamefont {Clerk}},\ and\
  \bibinfo {author} {\bibfnamefont {S.~M.}\ \bibnamefont {Girvin}},\ }``Quantum
  {Theory} of {Cavity}-{Assisted} {Sideband} {Cooling} of {Mechanical}
  {Motion}'',\ \bibfield  {journal} {\bibinfo  {journal} {Physical Review
  Letters}\ }\textbf {\bibinfo {volume} {99}},\ \bibinfo{pages} {093902} \href
  {https://doi.org/10.1103/PhysRevLett.99.093902}
(\bibinfo {year} {2007})\BibitemShut
  {NoStop}%
\bibitem [{\citenamefont {Safavi-Naeini}\ \emph {et~al.}(2013)\citenamefont
  {Safavi-Naeini}, \citenamefont {Chan}, \citenamefont {Hill}, \citenamefont
  {Gr{\"o}blacher}, \citenamefont {Miao}, \citenamefont {Chen}, \citenamefont
  {Aspelmeyer},\ and\ \citenamefont {Painter}}]{Ssafavi-naeini_laser_2013}%
  \BibitemOpen
  \bibfield  {author} {\bibinfo {author} {\bibfnamefont {A.~H.}\ \bibnamefont
  {Safavi-Naeini}}, \bibinfo {author} {\bibfnamefont {J.}~\bibnamefont {Chan}},
  \bibinfo {author} {\bibfnamefont {J.~T.}\ \bibnamefont {Hill}}, \bibinfo
  {author} {\bibfnamefont {S.}~\bibnamefont {Gr{\"o}blacher}}, \bibinfo
  {author} {\bibfnamefont {H.}~\bibnamefont {Miao}}, \bibinfo {author}
  {\bibfnamefont {Y.}~\bibnamefont {Chen}}, \bibinfo {author} {\bibfnamefont
  {M.}~\bibnamefont {Aspelmeyer}},\ and\ \bibinfo {author} {\bibfnamefont
  {O.}~\bibnamefont {Painter}},\ }``Laser noise in cavity-optomechanical
  cooling and thermometry'',\ \href
  {https://doi.org/10.1088/1367-2630/15/3/035007} {\bibfield  {journal}
  {\bibinfo  {journal} {New Journal of Physics}\ }\textbf {\bibinfo {volume}
  {15}},\ \bibinfo {pages} {035007} (\bibinfo {year} {2013})}\BibitemShut
  {NoStop}%
\bibitem [{\citenamefont {Nation}\ \emph {et~al.}(2008)\citenamefont {Nation},
  \citenamefont {Blencowe},\ and\ \citenamefont {Buks}}]{Snation_quantum_2008}%
  \BibitemOpen
  \bibfield  {author} {\bibinfo {author} {\bibfnamefont {P.~D.}\ \bibnamefont
  {Nation}}, \bibinfo {author} {\bibfnamefont {M.~P.}\ \bibnamefont
  {Blencowe}},\ and\ \bibinfo {author} {\bibfnamefont {E.}~\bibnamefont
  {Buks}},\ }``Quantum analysis of a nonlinear microwave cavity-embedded dc
  {SQUID} displacement detector'',\ \href
  {https://doi.org/10.1103/PhysRevB.78.104516} {\bibfield  {journal} {\bibinfo
  {journal} {Physical Review B}\ }\textbf {\bibinfo {volume} {78}},\ \bibinfo
  {pages} {104516} (\bibinfo {year} {2008})}\BibitemShut {NoStop}%
\bibitem [{\citenamefont {Yuan}\ \emph
  {et~al.}(2015{\natexlab{b}})\citenamefont {Yuan}, \citenamefont {Cohen},\
  and\ \citenamefont {Steele}}]{Syuan_silicon_2015}%
  \BibitemOpen
  \bibfield  {author} {\bibinfo {author} {\bibfnamefont {M.}~\bibnamefont
  {Yuan}}, \bibinfo {author} {\bibfnamefont {M.~A.}\ \bibnamefont {Cohen}},\
  and\ \bibinfo {author} {\bibfnamefont {G.~A.}\ \bibnamefont {Steele}},\
  }``Silicon nitride membrane resonators at millikelvin temperatures with
  quality factors exceeding 108'',\ \href {https://doi.org/10.1063/1.4938747}
  {\bibfield  {journal} {\bibinfo  {journal} {Applied Physics Letters}\
  }\textbf {\bibinfo {volume} {107}},\ \bibinfo {pages} {263501} (\bibinfo
  {year} {2015}{\natexlab{b}})}\BibitemShut {NoStop}%
\bibitem [{\citenamefont {Agrawal}\ and\ \citenamefont
  {Carmichael}(1979)}]{Sagrawal_optical_1979}%
  \BibitemOpen
  \bibfield  {author} {\bibinfo {author} {\bibfnamefont {G.~P.}\ \bibnamefont
  {Agrawal}}\ and\ \bibinfo {author} {\bibfnamefont {H.~J.}\ \bibnamefont
  {Carmichael}},\ }``Optical bistability through nonlinear dispersion and
  absorption'',\ \href {https://doi.org/10.1103/PhysRevA.19.2074} {\bibfield
  {journal} {\bibinfo  {journal} {Physical Review A}\ }\textbf {\bibinfo
  {volume} {19}},\ \bibinfo {pages} {2074} (\bibinfo {year}
  {1979})}\BibitemShut {NoStop}%
\end{thebibliography}
\end{document}